\documentclass[twocolumn,lineno,dvipsnames, adjustbox,usenames]{aastex63}
\usepackage{amsmath}
\usepackage{graphicx}
\usepackage[caption=false]{subfig}
\usepackage{placeins}
\usepackage{color}
\usepackage{dcolumn}
\usepackage{tensor}
\usepackage{bm}
\usepackage{microtype}
\usepackage{etoolbox}
\usepackage{amssymb}
\usepackage{mathrsfs}
\usepackage{accents}
\usepackage[normalem]{ulem}
\usepackage{xcolor}
\usepackage{hyperref}
\usepackage[all]{hypcap}
\usepackage[inline]{enumitem}
\usepackage[utf8]{inputenc}
\usepackage{xspace}
\usepackage{aas_macros}

\AtBeginDocument{\usepackage{booktabs}}

\newcommand{\beq}{\begin{equation}}
\newcommand{\eeq}{\end{equation}}

\newcommand{\Msun}{\ensuremath{\mathrm{M}_{\odot}}\xspace}

\newcommand{\nG}{\ensuremath{\dot{n}_{\rm G}}\xspace}
\newcommand{\nF}{\ensuremath{\dot{n}_{\rm F}}\xspace}
\newcommand{\nIII}{\ensuremath{\dot{n}_{\rm III}}\xspace}
\newcommand{\fG}{\ensuremath{f_{\rm G}}\xspace}
\newcommand{\fF}{\ensuremath{f_{\rm F}}\xspace}
\newcommand{\fIII}{\ensuremath{f_{\rm III}}\xspace}
\newcommand{\fGtoMain}{\ensuremath{\tilde{f}_{\rm G}}\xspace}
\newcommand{\nz}{\ensuremath{\dot{n}}\xspace}
\newcommand{\dNdz}{\ensuremath{dR/dz}\xspace}
\newcommand{\dNdzDetFrame}{\ensuremath{\frac{d R^d}{dz}}\xspace}
\newcommand{\dNdzDetFramekth}{\ensuremath{\frac{d R^d_k}{dz}}\xspace}
\newcommand{\dNdzSrcFrame}{\ensuremath{\frac{d R}{dz}}\xspace}
\newcommand{\Ndot}{\ensuremath{R}\xspace}
\newcommand{\NdotDet}{\ensuremath{R^d}\xspace}
\newcommand{\alphaF}{\ensuremath{\alpha_{\rm F}}\xspace}
\newcommand{\betaF}{\ensuremath{\beta_{\rm F}}\xspace}
\newcommand{\CF}{\ensuremath{C_{\rm F}}\xspace}
\newcommand{\muG}{\ensuremath{\mu_{\rm G}}\xspace}
\newcommand{\sigmaG}{\ensuremath{\sigma_{\rm G}}\xspace}
\newcommand{\tG}{\ensuremath{t_{\rm G}}\xspace}
\newcommand{\aIII}{\ensuremath{a_{\rm III}}\xspace}
\newcommand{\bIII}{\ensuremath{b_{\rm III}}\xspace}
\newcommand{\zIII}{\ensuremath{z_{\rm III}}\xspace}
\newcommand{\Npeak}{\ensuremath{N_{\rm peak}}\xspace}

\newcommand{\Nobs}{\ensuremath{N_{\rm obs}}\xspace}
\newcommand{\nbin}{\ensuremath{30}\xspace}

\newcommand{\nn}{\nonumber}

\newcommand{\ligomit}{LIGO Laboratory, Massachusetts Institute of Technology, Cambridge, Massachusetts 02139, USA}
\newcommand{\mki}{Kavli Institute for Astrophysics and Space Research, Massachusetts Institute of Technology, Cambridge, Massachusetts 02139, USA}

\newtoggle{acrotweak}
\togglefalse{acrotweak}

\ifdefined\acrotweak
    \toggletrue{acrotweak}
\fi

\iftoggle{acrotweak}{
  
}{
}

\newtoggle{commentsoff}
\togglefalse{commentsoff}

\ifdefined\nocomments
    \toggletrue{commentsoff}
\fi

\iftoggle{commentsoff}{
  \newcommand*{\kn}[1]{}
  \newcommand*{\wf}[1]{}
  \newcommand*{\cr}[1]{}
  \newcommand*{\sv}[1]{}
  \newcommand*{\comment}[1]{}
  
  \newcommand*{\todo}[1]{}
  \newcommand*{\warn}[1]{}

}{
  \newcommand*{\kn}[1]{\textsf{\color{blue} [\textbf{KEN}: #1]}}
  \newcommand*{\wf}[1]{\textsf{\color{magenta} [\textbf{WILL}: #1]}}
  
  \newcommand*{\sv}[1]{\textsf{\textcolor{orange}{\textbf{SALVO}: #1}}}
  
  \newcommand*{\comment}[1]{\textsf{\color{blue} [\textbf{NOTE}: #1]}}
  \newcommand*{\warn}[1]{\textsf{\color{red} [\textbf{WARNING}: #1]}}
  \newcommand*{\todo}[1]{\textsf{\color{red} [\textbf{TODO}: #1]}}

}

\graphicspath{{./}}

\begin{document}

\title{Probing multiple populations of compact binaries with third-generation gravitational-wave detectors}



\author{Ken K.~Y. Ng}
\email{kenkyng@mit.edu}
\affiliation{\ligomit}%
\affiliation{\mki}%

\author{Salvatore Vitale}
\affiliation{\ligomit}%
\affiliation{\mki}%

\author{Will M. Farr}
\affiliation{Department of Physics and Astronomy, Stony Brook University, Stony Brook, New York, 11794, USA}
\affiliation{Center for Computational Astrophysics, Flatiron Institute, 162 Fifth Avenue, New York 10010, USA}
\author{Carl L.~Rodriguez}
\affiliation{McWilliams Center for Cosmology, Department of Physics, Carnegie Mellon University, 5000 Forbes Ave, Pittsburgh, Pennsylvania 15213, USA}

\hypersetup{pdfauthor={Ng, Vitale, Farr, Rodriguez}}

\date{\today}

\begin{abstract}
Third-generation (3G) gravitational-wave (GW) detectors will be able to observe binary-black-hole mergers (BBHs) up to redshift of $\sim 30$. This gives unprecedented access to the formation and evolution of BBHs throughout cosmic history.
In this paper we consider three subpopulations of BBHs originating from the different evolutionary channels: isolated formation in galactic fields, dynamical formation in globular clusters and mergers of black holes formed from Population~III (Pop~III) stars at very high redshift.
Using input from populations synthesis analyses, we create two months of simulated data of a network of 3G detectors made of two Cosmic Explorers and one Einstein Telescope, consisting of $\sim16000$ field and cluster BBHs as well as $\sim400$ Pop~III BBHs.
First, we show how one can use a non-parametric model to infer the existence and characteristics of a primary and secondary peak in the merger rate distribution as a function of redshift.
In particular, the location and the height of the secondary peak around $z\approx 12$, arising from the merger of Pop~III remnants, can be constrained at $\mathcal{O}(10\%)$ level (95\% credible interval).
Then we perform a modeled analysis, using phenomenological templates for the merger rates of the three subpopulations, and extract the branching ratios and the characteristic parameters of the merger rate densities of the individual formation channels.
With this modeled method, the uncertainty on the measurement of the fraction of Pop~III BBHs can be improved to $\lesssim 10\%$, while the ratio between field and cluster BBHs can be measured with an uncertainty of $\sim 100\%$.
\end{abstract}

\section{Introduction}

Advanced gravitational-wave (GW) detectors such as LIGO~\citep{TheLIGOScientific:2014jea}, Virgo~\citep{TheVirgo:2014hva} and Kagra~\citep{Aso:2013eba} have dramatically increased our ability to study stellar mass black holes and the environments in which they form. The latest catalog released by the LIGO-Virgo-Kagra (LVK) collaboration includes 39 new GW detections, most of which are binary black holes (BBHs), bringing the total number of stellar mass black holes detected with GWs to over 100~\citep{GWTC1,GWTC2}.
These observations have already allowed for interesting astrophysical measurements~\citep{GWTC1,GWTC1rate,GWTC2,GWTC2rate}, such as hints for multiple formation channels~\citep{Farr:2017uvj,Zevin:2017evb,Belczynski:2017gds,Wong:2020ise,Zevin:2020gbd,Callister:2020vyz,Antonini:2020xnd}, hierarchical mergers~\citep{Fishbach:2017dwv,Gerosa:2017kvu,Doctor:2019ruh,Kimball:2019mfs,Kimball:2020qyd,Gerosa:2020bjb,Tiwari:2020otp}, as well as constraints on stellar physics~\citep{Farmer:2020xne,Fragione:2020miv,Bavera:2020uch}, primordial BHs~\citep{AliHaimoud:2017rtz,Bird:2016dcv,Sasaki:2016jop,Wong:2020yig,Hall:2020daa,Boehm:2020jwd,Hutsi:2020sol} and ultralight bosons~\citep{Arvanitaki:2016qwi,Brito:2017zvb,Ng:2019jsx,Ng:2020ruv}
Therefore, the growing set of BBHs enable studying both the properties of individual sources, as well as those of the underlying populations.

One of the key questions that can be addressed by GW astrophysics is how many such populations exist, and what are their characteristics. Multiple approaches have been proposed to address these questions, which ultimately rely on looking for features that would be expected in the BBH generated by each channel, such as their mass and spin distribution~\citep{Vitale:2015tea,Stevenson:2015bqa,Farr:2017gtv,Farr:2017uvj,Fishbach:2017zga,Fishbach:2019ckx,Kimball:2020opk,Talbot:2017yur,Talbot:2018cva,Doctor:2019ruh,Bouffanais:2019nrw,Miller:2020zox,Safarzadeh:2020mlb,Fishbach:2020qag}, eccentricity distribution~\citep{Lower:2018seu} or redshift distribution~\citep{Fishbach:2018edt,Callister:2020arv}.
However, due to the limited sensitivity of current GW detectors, the observed BBHs are relatively ``local'', with redshift $z\lesssim1$~\citep{GWTC1,GWTC1rate,GWTC2,GWTC2rate,IASO1O2rate,IASO2}.
Even as LIGO, Virgo and Kagra improve their sensitivities with the implementation of frequency dependent quantum squeezing~\citep{McCuller:2020yhw}, and better low-frequency isolation~\citep{Yu:2017zgi}, the detectors horizon will reach $z\sim 3$ only for the heaviest systems~\citep{Martynov:2016fzi,Hall:2019xmm}.

Second-generation detectors will therefore be unable to access mergers at redshifts larger than a few. This likely precludes the possibility of detecting binaries whose component black holes originated directly from the Population III (Pop~III) stars, whose formation peak could be as high as $z\sim 10$. Primordial black holes~\citep{Kinugawa:2014zha,Kinugawa:2015nla,Belczynski:2016ieo,Hartwig:2016nde,Raidal:2017mfl,Raidal:2018bbj}, might also be out of reach for existing facilities~\footnote{We note that there are studies suggesting the heaviest BBH detected in the first part of LIGO/Virgo O3 run - GW190521~\citep{Abbott:2020tfl,Abbott:2020mjq} - may be a Pop~III BBH~\citep{Kinugawa:2020xws,Tanikawa:2020abs}.}.
This will not be the case in the era of third-generation (3G) GW detectors such as Cosmic Explorer (CE)~\citep{Evans:2016mbw,Reitze:2019iox} and Einstein Telescope (ET)~\citep{VanDenBroeck:2010vx,Punturo:2010zz,Maggiore:2019uih}. In fact, 3G detectors will have horizons up to $z\gtrsim30$ and enable accessing most of the BBHs throughout the cosmic history~\citep{Hall:2019xmm}. Therefore, both detections and non-detections of high redshift BBHs with 3G detectors can provide significant constraints on the properties of Pop~III remnants.
This is particularly important as the remnants of Pop~III stars might be the light seeds that lead the formation of supermassive black holes early in cosmic history~\citep{2020ARAA..58..257G}. Whereas the space-based gravitational-wave detector LISA~\citep{Audley:2017drz} and electromagnetic missions such as the X-ray observatory Lynx~\citep{lynx} might reveal the presence of heavy (i.e. $M\gtrsim 100$~\Msun) black holes at redshifts of 10, ground-based 3G detectors might very well be the only way to searching for smaller building blocks: stellar mass black holes.

Using the current \emph{local} BBH merger rate estimate, $\sim 25~\mathrm{Gpc}^{-3}\mathrm{yr}^{-1}$~\citep{GWTC2rate}, the total merger rate of the BBHs in the Universe is inferred to be $\sim 10000$ per month , if the merger rate density follows the same evolution as the star formation rate (SFR)~\citep{Regimbau:2016ike}.
Nearly all of these BBHs would be detectable by 3G detectors, and most of which will also have very high signal-to-noise ratio (SNR), leading to very precise distance (and hence redshift, assuming a known cosmology) measurements~\citep{Vitale:2016aso,Vitale:2016icu,Vitale:2018nif}.
The large number of loud observations thus allows inferring the morphology of merger rate densities, in both parametric and non-parametric ways~\citep{Broeck:2013rka,Vitale:2018yhm,Safarzadeh:2019pis,Romero-Shaw:2020siz}.
In ~\citet{Vitale:2018yhm}, we showed how combining redshift measurements of tens of thousands of GW observations in 3G detectors and assuming all BBHs are formed in galactic fields, one can infer the SFR history and the time-delay distribution.
In light of the properties of the BBHs detected by LIGO and Virgo in the last few years, it has become harder to assume that all black holes are formed in galactic fields, since many of the black holes being detected are consistent with having formed dynamically, in globular or nuclear clusters or in the disks of active galactic nuclei (AGN)~\citep{Bartos:2016dgn,Yi:2019rwo,Yang:2019cbr,Yang:2020lhq,Grobner:2020drr,Tagawa:2019osr,Tagawa:2020qll,Tagawa:2020dxe,Samsing:2020tda}. In this paper, we greatly extend our previous work~\citep{Vitale:2018yhm} and explore the possibility of simultaneously identifying and constraining the properties of \emph{multiple} formation channels. In particular, we simulate universes where black holes can be formed in galactic fields, globular clusters and from Pop~III stars. We show how well one can constrain the properties of each channel, and their branching ratios. In particular, we focus on the evidence for a high-redshift ($z\gtrsim 5$) population, which would be the smoking gun of formation outside of the traditional evolutionary pathways.

\section{Astrophysical models}\label{sec:pops}

In this section we summarize the main astrophysical properties of the BBH
formation channels that we consider in the paper.
Astrophysical BBHs are believed to form in various environments, such as binary stellar evolution in galactic fields~\citep{OShaughnessy:2016nny,Dominik:2012kk,Dominik:2013tma,Dominik:2014yma,deMink:2015yea,Belczynski:2016obo,Stevenson:2017tfq,Mapelli:2019bnp,Breivik:2019lmt,Bavera:2019fkg,Broekgaarden:2019qnw}, dynamical formation through multi-body interactions in star clusters (from low-mass stars to large nuclear star clusters)~\citep{2000ApJ...528L..17P,Antonini:2020xnd,Santoliquido:2020axb,Rodriguez:2015oxa,Rodriguez:2016kxx,Rodriguez:2018rmd,DiCarlo:2019pmf,Kremer:2020wtp,Rodriguez:2015oxa,Rodriguez:2016kxx,Rodriguez:2018rmd,Antonini:2020xnd} or AGN disk~\citep{Bartos:2016dgn,Yi:2019rwo,Yang:2019cbr,Yang:2020lhq,Grobner:2020drr,Tagawa:2019osr,Tagawa:2020qll,Tagawa:2020dxe,Samsing:2020tda}, from Population~III (Pop~III) stars~\citep{Kinugawa:2014zha,Kinugawa:2015nla,Hartwig:2016nde,Belczynski:2016ieo} or primordial black holes~\citep{Carr:1974nx,AliHaimoud:2017rtz,Clesse:2016vqa,Bird:2016dcv,Sasaki:2016jop,Raidal:2017mfl,Raidal:2018bbj,Wong:2020yig,Boehm:2020jwd,Hall:2020daa}.

For simplicity, in this paper we use galactic fields and globular cluster as the only main populations at low redshift, and Pop~III stars as the only high-redshift channel. The analysis can be easily extended to even more population, at the price of an increased computational cost.

Hereafter, we label the BBHs in the three formation channels as \textit{field binaries}, \textit{cluster binaries} and \textit{Pop~III binaries}.
Since an astrophysical BH is a remnant of stellar collapse, the merger rate history of each channel is correlated with the SFR and with the time delay from the binary formation to merger.
Cluster and field binaries consist of BH-remnants leftover from Pop I/II stars, whose corresponding SFR peaks at late times: $z\sim3$~\citep{Vangioni:2014axa,Madau:2014bja}. Accounting for the typical time delay between binary formation and merger (${\sim}~10$~Myr to ${\sim}~10$~Gyr), their merger rates are expected to peak at around $z \sim 2$~\citep{Dominik:2013tma,Belczynski:2016obo,Mapelli:2019bnp,Rodriguez:2018rmd}.

Pop~III stars are instead formed at early times, $z~{\gtrsim}~10$, from primordial gas clouds at extremely low metallicity~\citep{Vangioni:2014axa,deSouza:2011ea}. The ``metal-free'' environment reduces the stellar wind mass loss during the binary evolution~\citep{Baraffe:2000dp} so that Pop~III stars might be more massive than later stellar populations. Eventually, heavy BH remnants are left behind that merge in a short timescale, resulting on a merger rate density that could peak at around $z\sim10$~\citep{Belczynski:2016obo,Hartwig:2016nde,Kinugawa:2014zha,Kinugawa:2015nla}.

The fact that different formation channels result in different merger rate distributions as a function of redshift, especially for Pop~III remnants, can be exploited to infer properties of BBH populations solely based on the redshift distribution of the detected sources.
More elaborate tests can be envisaged, that also rely on other distinguishing features, e.g. masses, spins or eccentricity of the sources. In this study we will show that tests based on the redshift distribution alone can already provide significant constraints, while also having the benefit of being model-independent, at least in some of the implementations we demonstrate below. This seems particularly desirable, since the true distribution of intrinsic parameters such as masses and spins is highly uncertain, especially for Pop~III remnants.

\begin{figure}[h]
\includegraphics[width=0.9\columnwidth]{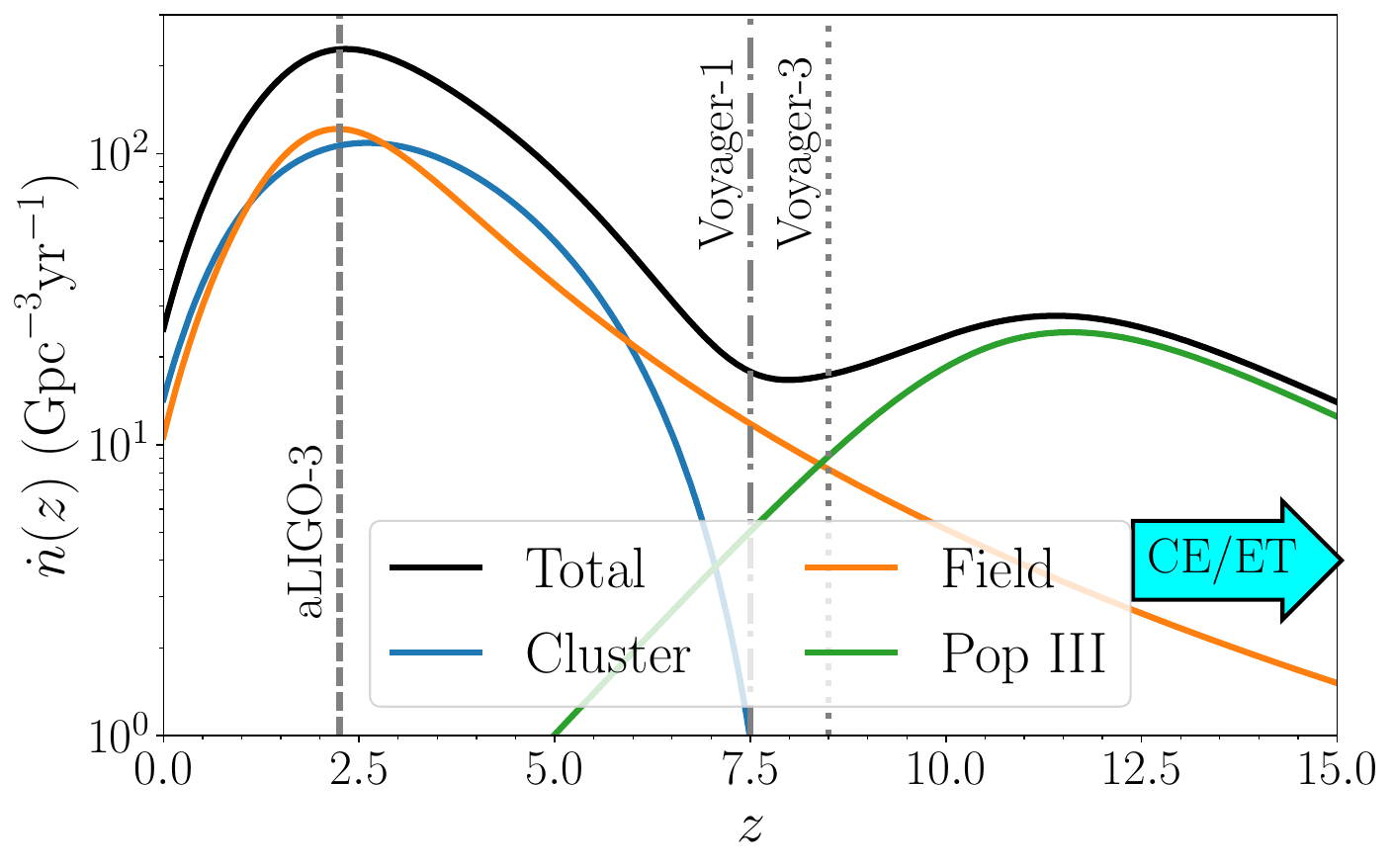}
\caption{\label{fig:horizon}
The merger rate densities of field (orange), cluster (blue) and Pop~III (green) binaries, together with the overall merger rate (black), given by the sum of the three populations.
For the purposes of this plot, field and cluster merger rate densities are normalized such that they produce the same number of binaries up to $z$ of $15$.
The Pop~III merger rate density is scaled so that  its peak has an amplitude of $1/10$ relative to the peak of the field and cluster merger rate densities.
The vertical lines indicate the detector horizons, $z_h$, to BBHs with total masses (in the comoving frame) $\leq100\Msun$, in a network of three advanced detectors $(z_h \sim 2.25)$, a single Voyager~\citep{Adhikari:2019zpy} $(z_h \sim 8.5)$ and a network of three Voyager-like detectors $(z_h \sim 12.1)$, respectively~\citep{Hall:2019xmm}.
The SNR threshold for detection is set to be 8 for a single detector and 12 for a detector network.
}
\end{figure}
In Fig.~\ref{fig:horizon}  we show the ``true'' merger rate densities of three formation channels we use in this study, and focus on key features of their shapes (we will discuss later the branching ratios, i.e. the relative scale).
The rate densities of field (orange), cluster (blue) and Pop~III (green) binaries are phenomenological fits (details in Appendix~\ref{sec:phenom}, Eqs.~\eqref{eq:field},~\eqref{eq:cluster}~and~\eqref{eq:pop3}) of the population synthesis simulation from~\cite{Belczynski:2016obo,Rodriguez:2018rmd,Belczynski:2016ieo}, respectively.
We notice that the high-redshift tail of the cluster merger rate density is much steeper than that of the field binaries.
This is largely due to the choice of model: the cluster merger rates are based on the model of globular cluster formation from \citep{2019MNRAS.482.4528E}, which goes to zero at $z\sim 10$.  That, combined with the delay between cluster formation and BBH mergers (since BHs in clusters can only merge after the cluster has formed \emph{and} the BHs have sunk to the center due to dynamical friction, a process which can take $\sim 100$Myr \cite[e.g.,][]{2015ApJ...800....9M}) causes a steeper slope in the merger rate at high $z$.
The field and cluster merger rate densities peak at similar values, $z\sim 2.2$ and $z\sim 2.6$, respectively, whereas the Pop~III merger rate density peaks much later, at  $z\sim 11.6$.

The vertical lines in Fig.~\ref{fig:horizon} report the horizon of future ground-based detector networks.
Advanced detector networks can observe BBHs up  to the low-redshift peak of the merger rate densities of the two dominating channels, field and clusters.
However, as the plot shows, for $z\lesssim2$, the merger rate densities of both field and cluster binaries are quite similar.
Hence  advanced detectors are unlikely to be able to disentangle the two channels using only redshift information (as mentioned above, one can use other features, at the price of making the analysis more model-dependent).
The situation improves with a single Voyager detector (``Voyager-1'' line), which can access most of the field and cluster binaries up to $z\lesssim 8$ and therefore exploit the expected difference in their merger rate after the peak to characterize the two channels.
A network of 3 Voyager-like detectors (``Voyager-3'' line) can extend the horizon to a redshift where the contribution to the total merger rate of the field and the Pop~III channel might become comparable. However, it is only with 3G detectors that one can access the peak of the merger rate from Pop~III. In fact, the horizon of CE and ET to heavy BBHs is outside of the range of Fig.~\ref{fig:horizon} (as indicated by the cyan arrow in the bottom right corner), at $z\sim100$ .
As we will show in the following sections the fact that the horizon of 3G detectors extends well beyond the expected peak of Pop~III mergers  allows for both modeled and unmodeled tests of the existence of such subpopulation.

\section{Results}\label{sec:results}

In this work, we follow two approaches to measure the comoving-frame merger rate density $\dNdz$: (i) a unmodeled approach that utilizes Gaussian process regression (GPR) to infer $\dNdz$ as a piecewise function over several redshift bins~\citep{Mandel:2016prl}; and (ii) a modeled approach in which we use phenomenological models for the various subpopulations~\citep{Farr:2013yna,Vitale:2018yhm}.
In both cases, we use hierarchical a Bayesian inference framework~\citep{Farr:2013yna,Mandel:2018mve,Thrane:2018qnx,Wysocki:2018mpo,Vitale:2020aaz} to measure the parameters of the population(s). More details are provided in Appendix~\ref{sec:stats}. 
Details about the implementation of the GPR analysis can be found in Appendix ~\ref{sec:gpr}, whereas Appendix~\ref{sec:phenom} reports the functional forms of the modeled populations. The priors used in the analysis are documented in Appendix~\ref{app:hyperprior}.

To study how well the models can identify the Pop~III subpopulation, we perform a mock-data challenge by simulating 18 different universes, which contain two-months worth of BBH data with a majority of cluster or field binaries. The detailed setup of the simulations can be found in Appendix~\ref{sec:injection}.

In the following, we will focus on the measurement of the  \emph{volumetric} merger rate density, $\dot{n}(z)\equiv d\Ndot/dV_c$, rather than $\Ndot$ itself (We will use an index to indicate the volumetric merger rate in a specific channel, e.g. $\dot{n}_{\rm{III}}(z)$ for the volumetric merger rate in the Pop.~III channel. ``F'' will indicate the field channel and ``G'' the globular cluster channel).
We will also report branching ratios between the channels. $f_{\rm{III}}$ will indicate the fraction of Pop.~III mergers over the \emph{total}, whereas $\tilde{f}_{G}$ will indicate the fraction of cluster binaries over the sum of field and cluster binaries. 

Our modeled approach naturally provides more information about the characteristic parameters of each channel, and their correlations. Those are discussed in Appendix~\ref{app.Hyperpost}.

\subsection{Unmodeled analysis}

\begin{figure*}[ht]
\centering
\subfloat[$\fIII=0$\label{fig:GPRPosPopIII}]{\includegraphics[width=0.9\columnwidth]{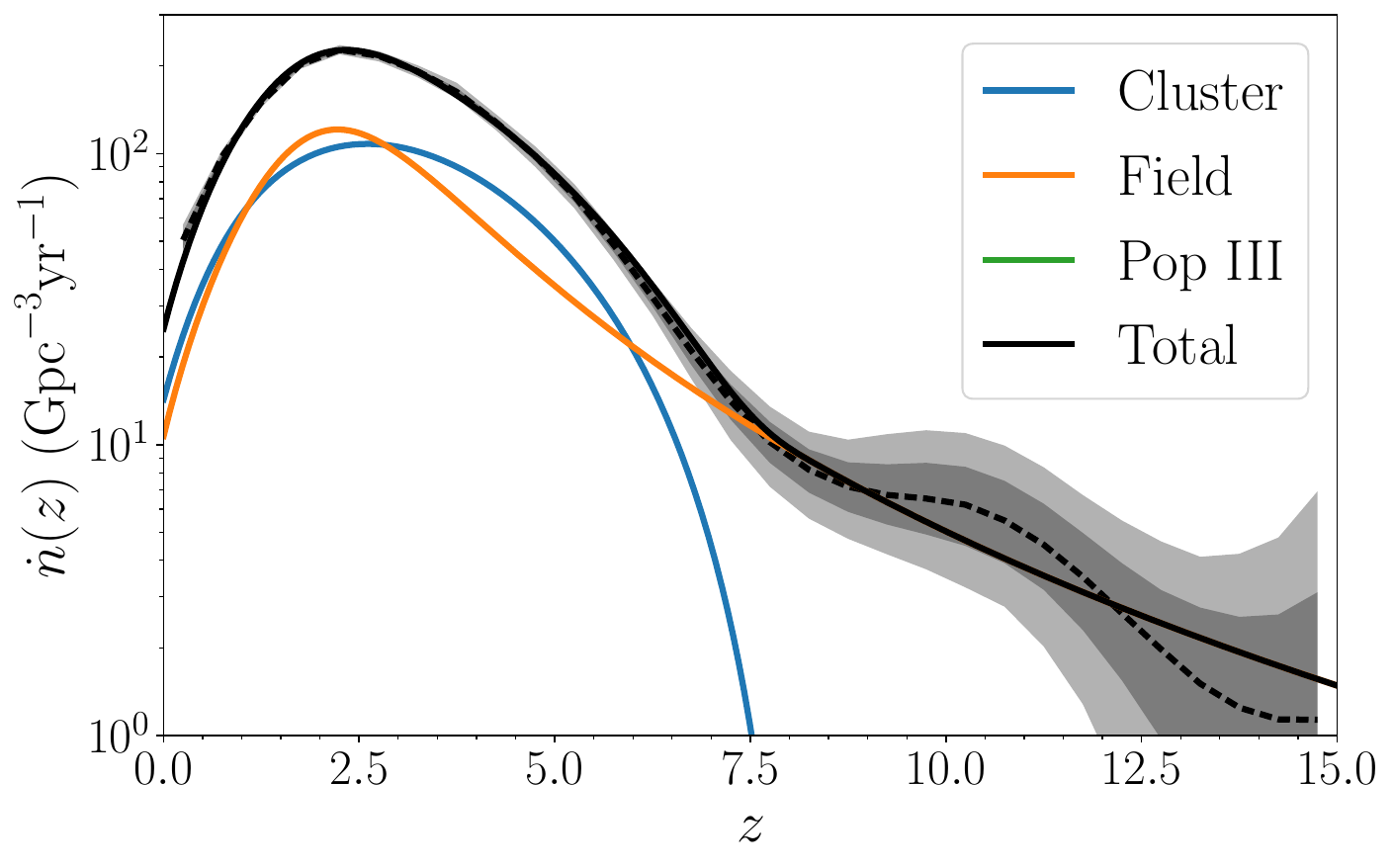}}
	\subfloat[$\fIII=0.024$]{\includegraphics[width=0.9\columnwidth]{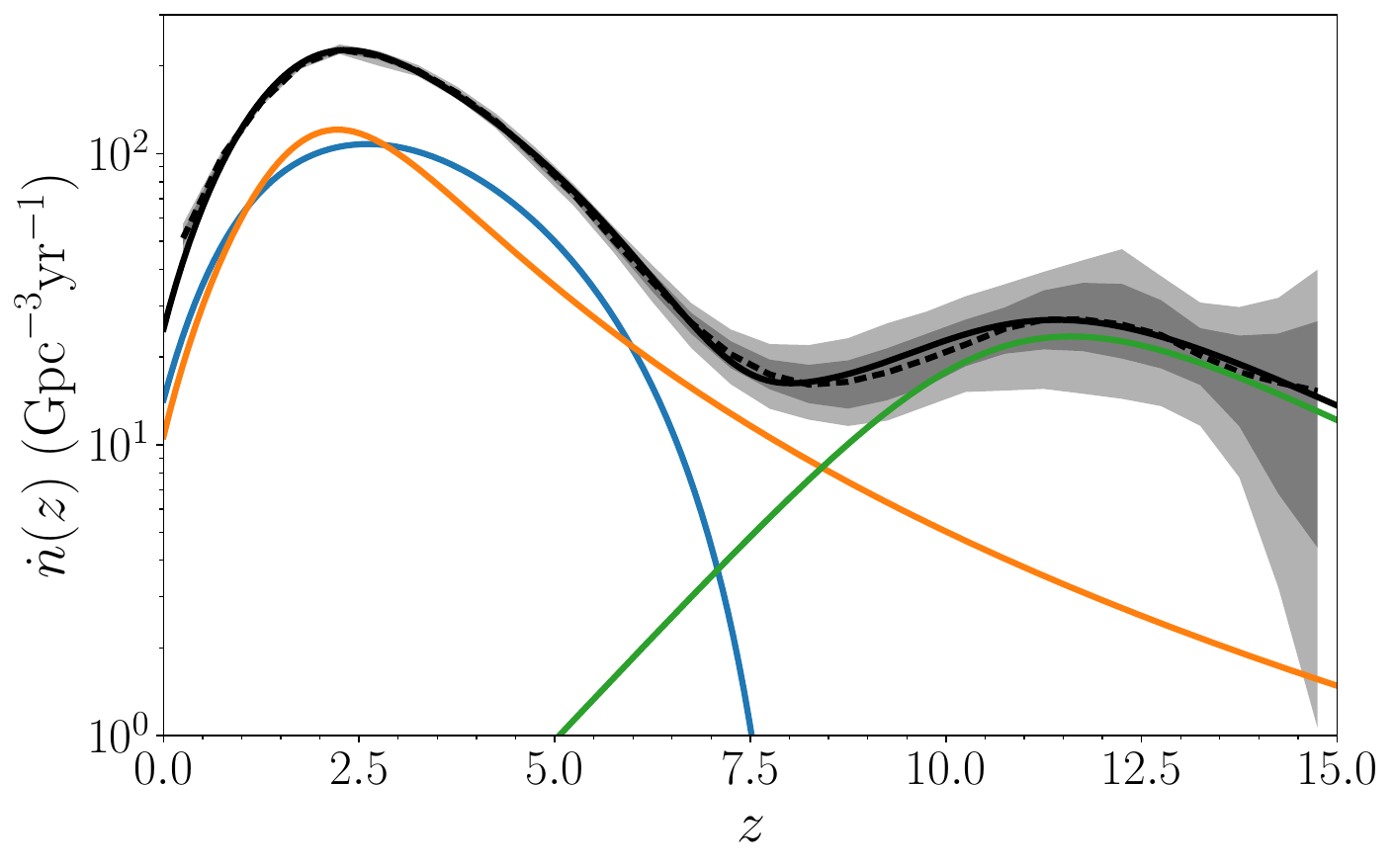}}
\caption{
Reconstruction of $\nz$ for the $\fGtoMain=0.5$ universes with (right panel) and without (left panel) Pop III binaries using GPR.
The grey color band and dashed line shows the 68\% (darker color) and 95\% (lighter color) credible intervals, and the median of recovered $\nz$, respectively.
The black, orange, blue and green solid lines are the fiducial $\nz$, $\nF$, $\nG$ and $\nIII$, respectively.
Since we cannot model each subpopulation with the nonparametric approach, no hyper-posterior can be drawn for each branch.
}
\label{fig:GPRPos}
\end{figure*}

Figure~\ref{fig:GPRPos} shows our inference on the volumetric merger rate $\nz$ obtained with the unmodeled GPR approach.
The black colored bands report the 68\% and 95\% credible intervals for the universes with (right panel) and without (left panel) Pop~III binaries. The colored lines show the true volumetric merger rate of the individual populations. For both of the panels the true branching ratio between field and cluster binaries is $\fGtoMain=0.5$. We find that the true $\nz$'s (black solid lines) lie within the 95\% credible intervals in both cases.
While the relative uncertainty on $\nz$ is at a percent level at $z\lesssim6$, it increases to $\mathcal{O}(100\%)$ at $z\gtrsim8$.
This is because (i) the SNR of each source decreases with the distance, and (ii) the number of sources in each redshift bin is decreasing as the differential comoving volume shrinks at earlier times in the history of the universe.

Perhaps the most attractive feature of the unmodeled analysis, is that we can find some evidence for the presence of an high-redshift subpopulation, even without strong modeling.
The simplest way of doing this is to look for local peak(s) in $\nz$. At the very minimum, we would expect to find evidence for the ``main'' peak at $z\sim 2$, arising from the merger in fields and clusters, while high-redshift peaks would be indicative of a different subpopulation.

We implement a peak finder algorithm simply by asking that the first derivative of $\nz$ is zero and the second is negative: $d\dot{n}/dz=0$ and $d^2\dot{n}/dz^2<0$.
Some care is required to avoid false positives due to natural oscillations in the results of the GRP which are not due to astrophysical maxima (or minima) but only to the underlying Gaussian process. These are particularly visible  at high redshifts in Fig.~\subref*{fig:GPRPosPopIII}.
To mitigate the effect of these fluctuation, we require the height of any high redshift peaks to be at least $1/10$ of the height of the low redshift peak, as well as an intra-peak separation larger than $\Delta z = 1$. These requirements are based on the expected excess of Pop~III as discussed in Sec.~\ref{sec:injection} and arguably represent the only modeling involved in the GRP approach that we describe.

\begin{figure}[h]
\includegraphics[width=0.9\columnwidth]{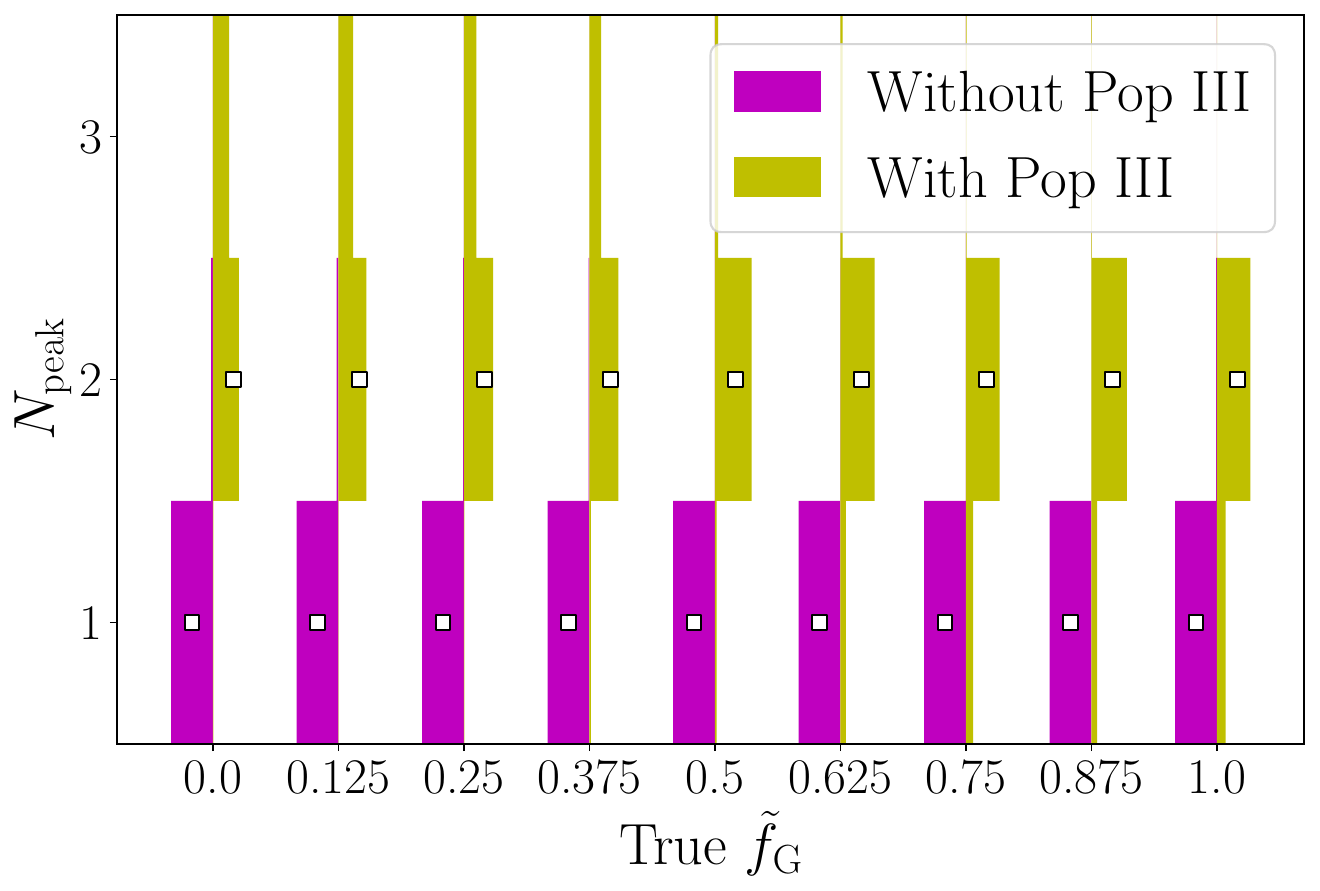}
\caption{\label{fig:Npeaks}
Posterior distributions of $\Npeak$ of $\nz$ for all 18 universes.
Since $\Npeak$ is a discrete measure, the distributions are represented by histograms.
For the universes with (yellow) and without (purple) Pop~III binaries, the true $\Npeak$ are 2 and 1, respectively.
}
\end{figure}

With these two restrictions, we count the number of peaks $\Npeak$ for each $\nz$ sample of the GRP in every simulated universe. The results are shown in Fig.~\ref{fig:Npeaks} as a function of the true branching ratio cluster/field.
For the universes without Pop~III binaries, we recover a single peak, as expected, with $>99\%$ probability (purple histograms).
This implies the non-existence of a secondary peak whose relative height is $\geq10\%$ of the primary peak.
On the other hand, the true value for the number of peaks,  $\Npeak=2$, is found at $\gtrsim90\%$ credibility  for in the universes with Pop~III binaries (yellow histograms).

We observe that in the universes with Pop~III binaries the posterior on the number of peaks has a secondary mode at $\Npeak=3$ for $\fGtoMain\lesssim 0.5$. We explain this as follows: for $\fGtoMain\lesssim 0.5$, the field binaries are the dominating channel, and as clear in Fig.~\ref{fig:GPRPos} produce a flatter high-redshift tail than the cluster channel.
Hence it is easier to produce multiple peaks due to fluctuation and induces a leakage to $\Npeak\geq3$. On the other hand, when $\fGtoMain \gtrsim 0.5$, the cluster binaries are dominating and do not contribute to the merger rate at $z\gtrsim8$ where Pop~III population becomes the only source of BBHs, which makes the high-redshift peak narrower and hence easier to reveal.
But the Poisson fluctuation of the high-redshift bins in $\dNdz$ may still lead to an underestimation of the relative height below our $10\%$ threshold, inducing a small contamination at $\Npeak=1$.
We note that the above trend is subject to the model uncertainty of our chosen simulation data in the high-redshift region.

This method also allows measuring the location(s) of the peak(s), which would be useful to understand the population properties.
For instance, the shift of the primary peak relative to the star formation rate could inform the typical time delay to merger and a hint of metallicity evolution~\citep{Chruslinska:2018hrb,Santoliquido:2020axb}.
In addition, constraining the high redshift peak to $z\gtrsim8$ would provide support to the existence of Pop~III binaries.

\begin{figure}
\includegraphics[width=0.9\columnwidth]{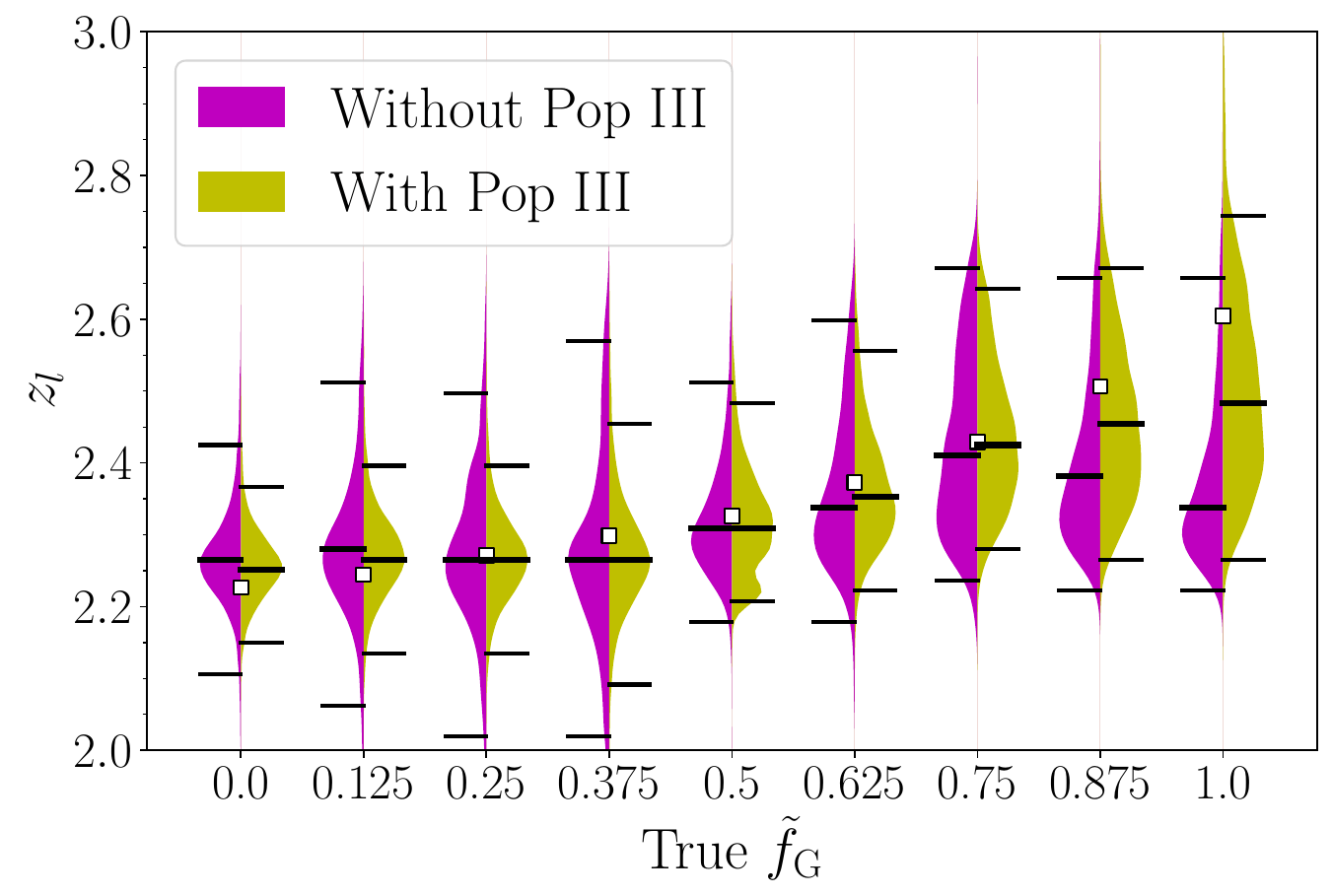}
\caption{\label{fig:firstPeaks}
Hyper-posterior of $z_l$ for all universes, split into two categories: with (yellow) and without (purple) Pop III binaries.
In each half-leaf, the upper and lower black dashes mark the 95\% credible interval, and the middle black dash locates the median.
The square markers indicate the true values of $z_l$.
}
\end{figure}
We first show the inferred distribution of the low redshift peak, $z_l$, for all universes in Fig.~\ref{fig:firstPeaks}.
All measurements of $z_l$ constrain the low redshift peaks to $z<3$, with 95\% credible-interval uncertainties of $\sim40\%$.
The true values are contained within the uncertainty.

\begin{figure}
\includegraphics[width=0.9\columnwidth]{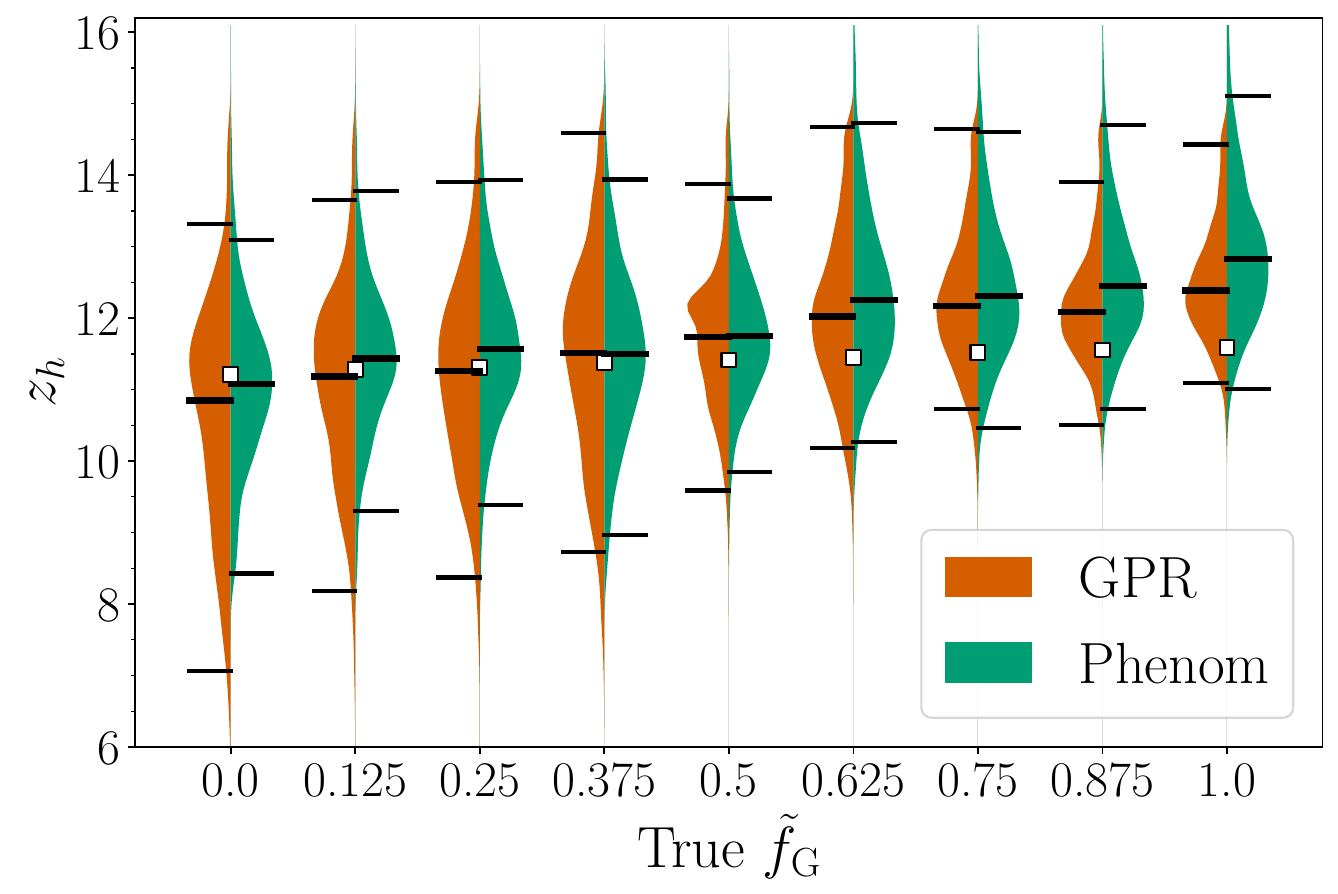}
\caption{\label{fig:secondPeaks}
Hyper-posterior of $z_h$ for the 9 universes with Pop~III binaries, conditioned with $\dot{n}(z_h)\geq0.1\dot{n}(z_l)$ inferred by the GPR model (orange) and the phenomenological (Phenom) model (green).
Other plot settings are the same as Fig.~\ref{fig:firstPeaks}.
}
\end{figure}
\begin{figure*}
\centering
	\subfloat[$\fIII=0$\label{fig:PMPosNull}]{\includegraphics[width=0.9\columnwidth]{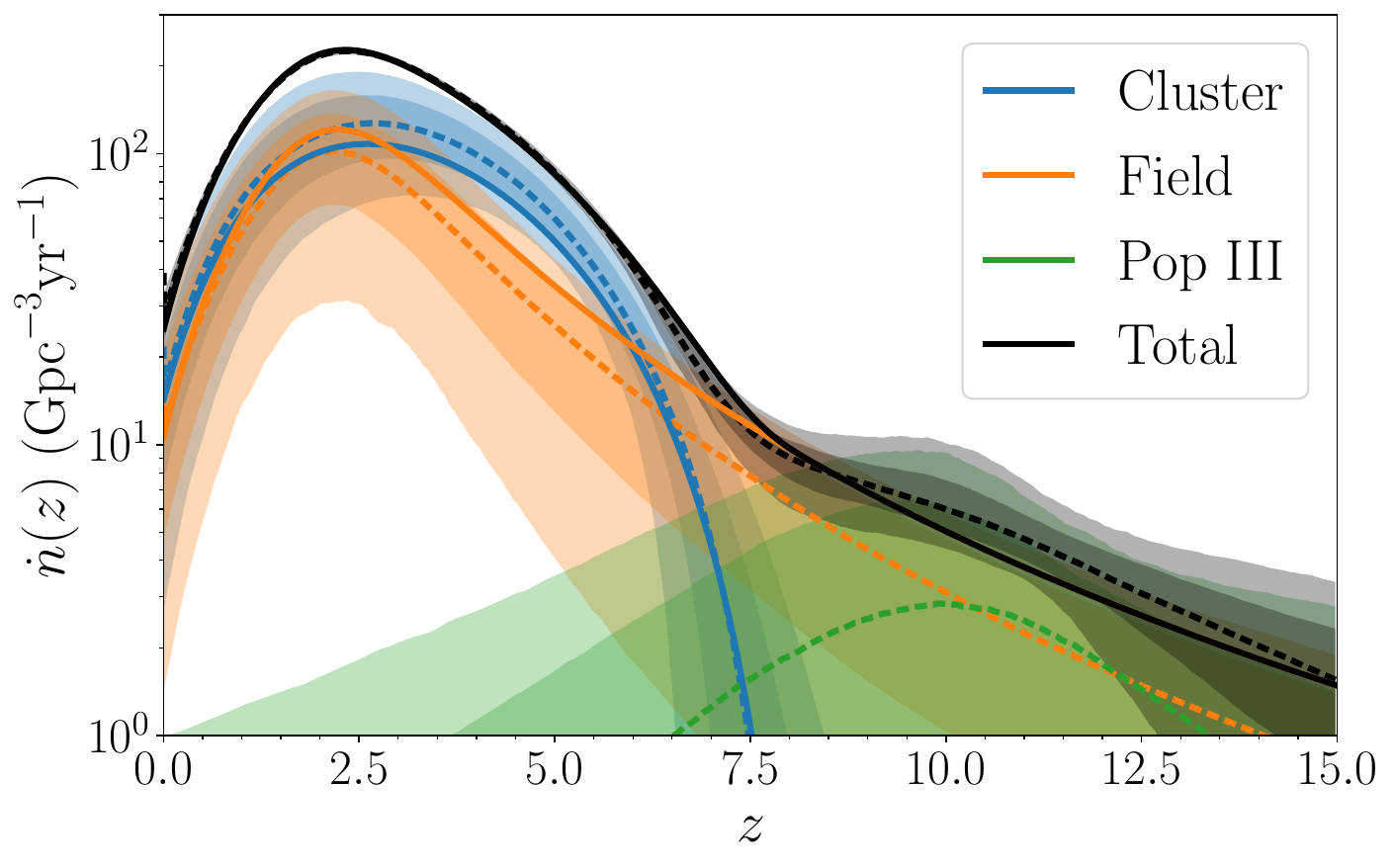}}
	\subfloat[$\fIII=0.024$\label{fig:PMPosPopIII}]{\includegraphics[width=0.9\columnwidth]{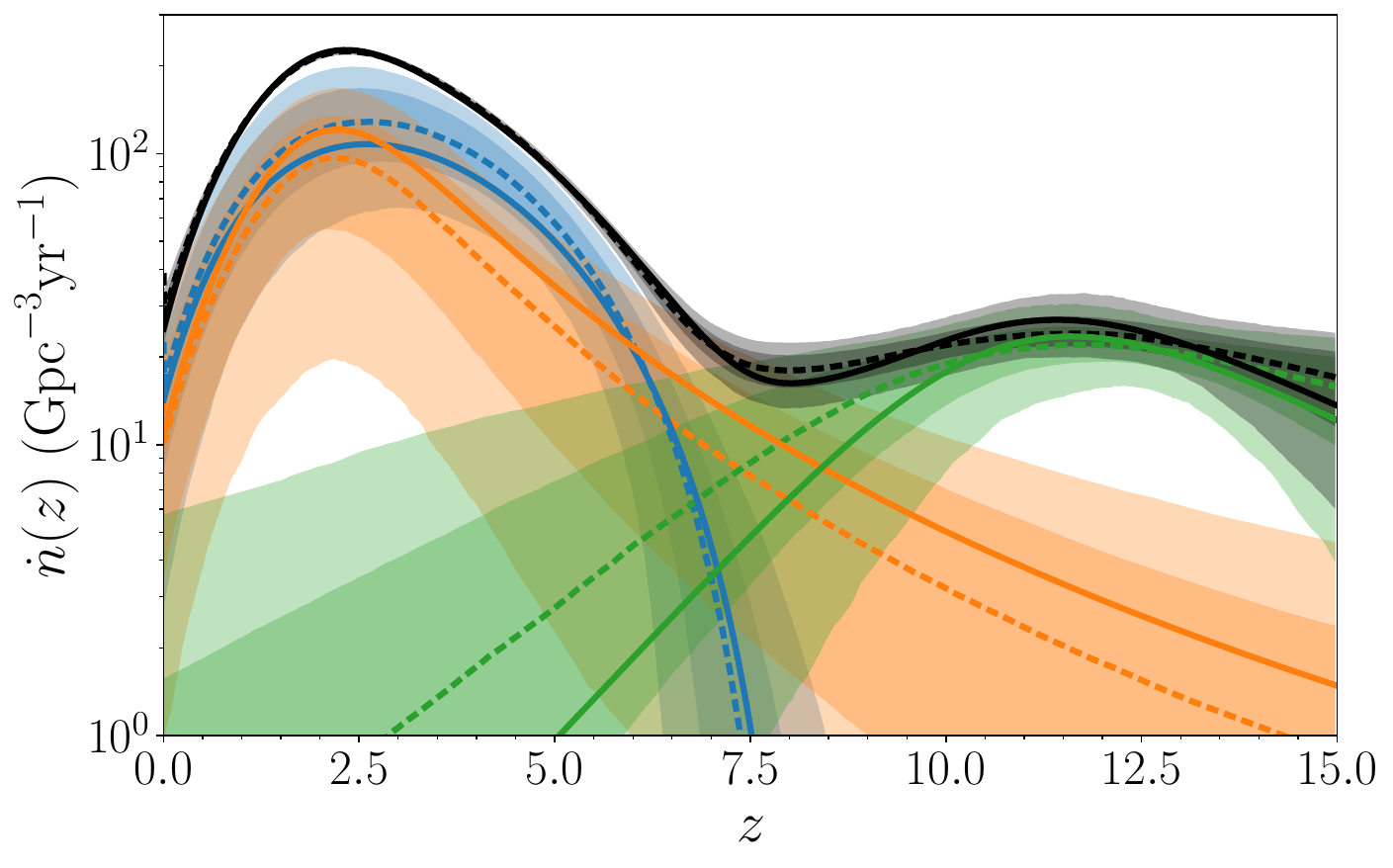}}
\caption{
Reconstruction of merger rate densities for the $\fGtoMain=0.5$ universes with (right panel) and without (left panel) Pop~III binaries using phenomenological models.
The orange, blue, green and black dashed lines are the medians of the recovered $\nF$, $\nG$, $\nIII$ and $\nz$, respectively.
The corresponding color bands are the 68\% (darker colors) and 95\% (lighter colors) credible intervals of the reconstruction.
The true merger rate densities are shown as the dashed lines of the same colors.
The green solid line is invisible since there are no Pop~III binaries in this universe.
}
\label{fig:PMPos}
\end{figure*}

Next, we look at the measurements of the high redshift peak's location $z_h$ for the universes with Pop~III binaries, as shown by the orange violins in Fig.~\ref{fig:secondPeaks}.
In all cases, the true value $\hat{z}_h=11.6$ lies within the 95\% credible intervals. The lower bound of the credible interval is above $z\sim7$ for all values of  $\fGtoMain$, indicating that one can confidently place the secondary peak at redshift much larger than where the star formation peaks. The widths of the credible intervals decrease from $\sim50\%$ to $\sim25\%$ when $\fGtoMain$ increases from 0 to 1. This may be again explained by the steeper redshift tail in the cluster population, which makes the Pop~III peak easier to resolve.

\subsection{Phenomenological analysis}

Having shown how a simple non-parametric model can already provide insight into the existence of a high-redshift population of BBHs, we now repeat the hierarchical inference using the phenomenological model described in Appendix~\ref{sec:phenom}.

We start by showing the posteriors on the peak of the merger rate density for the Pop~III BBHs, this time obtained as one of the parameters of the phenomenological Pop.~III model~\footnote{That is, $z_{\rm{III}}$ of the Pop.~III model described in Appendix~\ref{sec:phenom}}, in Fig.~\ref{fig:secondPeaks}, green violins. Remembering that the orange violins in the same plot reports the measurement we obtained with the GPR approach, we find that the two methods yield very consistent results, with the widths of 95\% credible intervals varying by $\sim 10\%$ at most.
The consistency between the two approaches highlights the promise of the nonparametric approach in revealing the existence and location of an high-redshift population.
However, the phenomenological model directly describes the morphology of each subpopulation and thus allows extracting information about each individual population, which cannot be accessed by the nonparametric approach.

In Fig.~\ref{fig:PMPos} we show the inferred $\nz$ for the simulated $\fGtoMain=0.5$ universes with (right panel) and without (left panel) Pop~III binaries~\footnote{These are the same two universes of Fig.~\ref{fig:GPRPos}}. For each population, the solid line represents the true merger rate, whereas the dashed line and the colored bands represents the median, and the 68\%/95\% credible intervals.

The total $\nz$ (black colored band) can be well constrained within a few percent level up to $z\sim6$.
For $\fIII=0$, the relative uncertainty rises to ${\sim}100\%$ at $z\gtrsim10$.
Even at low redshift, the uncertainty of $\nF$ (orange color band) and $\nG$ (blue color band) is about $50\%$, ${\sim}10$ times larger than that of the total rate, $\nz$.
This is because the morphology of $\nF$ and $\nG$ is similar at $z\lesssim3$, where most of the BBHs can be detected with a precise distance measurement.
Conversely, $\nF$ and $\nG$ are easier to distinguish at $z\gtrsim5$ where, as discussed before, $\nG$ is declining more rapidly than $\nF$, which instead has a long tail at $z\gtrsim 8$.
Overall, the similarity in the morphology of the low-redshift volumetric merger rate of the two dominating channels induces degeneracies, hence boosts uncertainties, in the the individual merger rates, $\nF$ and \nG.

Considering now the universe with $\fIII=0.024$ (right panel), we find that rate of Pop~III curve (green colored band) near its peak at $z \sim 12$ can be measured with a relative uncertainty of $\sim 50\%$, while the uncertainty of other channels remains similar to the universe with $\fIII=0$.

We compare the phenomenological recovery of the total $\nz$ to the GPR recovery (Fig.~\ref{fig:GPRPos}) for the same universes, and find that the typical uncertainty of $\nz$ is smaller by a factor of $\sim2$. This is not surprising, since the phenomenological approach uses models for the subpopulations, which inform the recovery of the overall merger rate.
Naturally, the price to pay for the improved precision is to have made the results depend on the goodness of the models.

\begin{figure}[h]
\includegraphics[width=0.9\columnwidth]{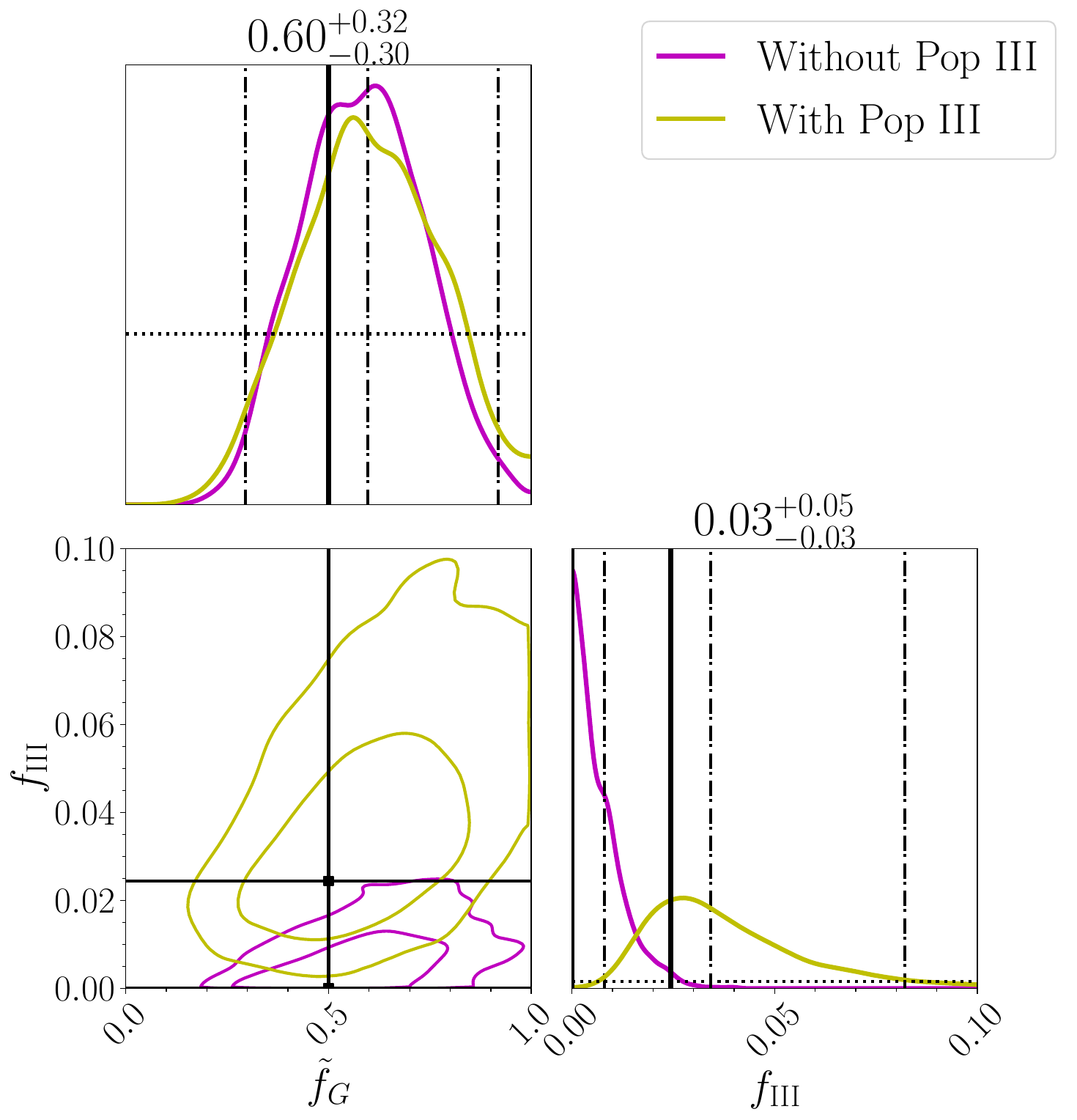}
\caption{\label{fig:fGCfIII}
Hyper-posterior of the population fractions, $(\fGtoMain,\fIII)$, for the universe with Pop~III binaries $(\fGtoMain, \fIII)=(0.5,0.024)$ (yellow).
For each marginalized 1D posterior (purple solid line in each diagonal slot), the left and right black dashed-dotted lines mark the 95\% highest posterior density credible interval, the middle black dashed-dotted line locates the median, and the black dotted line shows the prior.
The numerical values of median values and 95\% credible intervals are reported above the diagonal slots.
The off-diagonal slots show the marginalized 2D posteriors, with the contours representing the 68\% and 95\% credible intervals.
The black markers and solid black lines indicate the true values, which lie within the 68\% credible interval.
As a comparison, we overlay the same hyper-posterior for the universe without Pop~III binaries $(\fGtoMain, \fIII)=(0.5,0)$ (purple), whose 95\% credible intervals and median values are not shown.
}
\end{figure}

It is worth looking at the correlations between the hyper-parameters of the the various subpopulations. Some correlation should be expected since for example the number of sources at high redshift might be potentially explained by the model either with a larger fraction of field binaries, which have a fat high-redshift tail, or by binaries in the Pop~III channel.
In Fig.~\ref{fig:fGCfIII}, we show the marginalized 2D contours of the $(\fGtoMain, \fIII)$ pair for the universes $\fGtoMain=0.5$ with (yellow) and without (purple) Pop~III.
A positive correlation between the two parameters is clearly visible. This is caused by the partial model degeneracy between $\nF$ and $\nIII$.
This goes exactly in the direction one would expect: underestimating $\fIII$ means that the model must increase the number of field binaries to account at least partially for the high-redshift binaries. But if the number of field binaries increases, $\fGtoMain$ must decrease.

\begin{figure}[h]
\includegraphics[width=0.9\columnwidth]{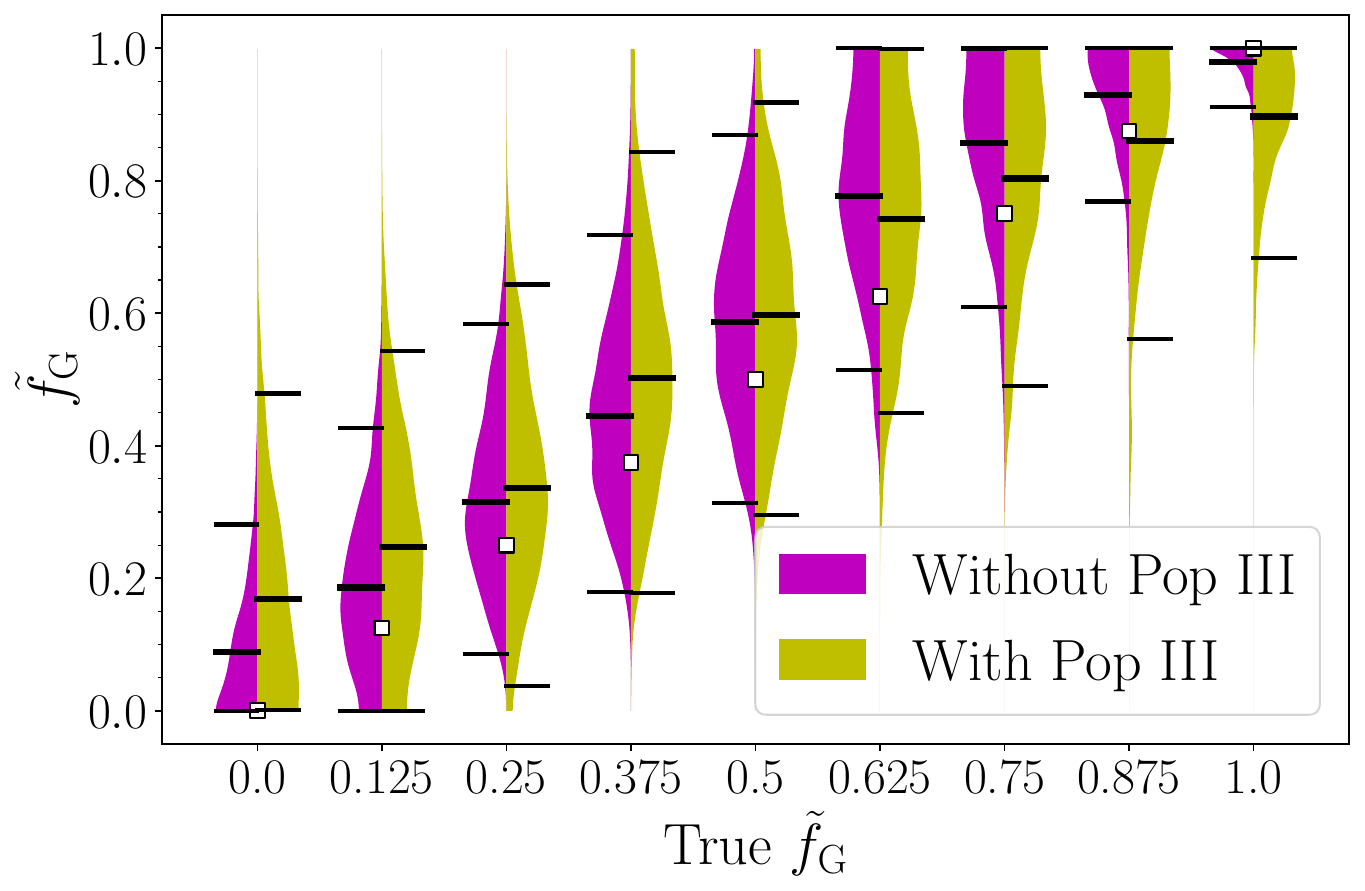}
\caption{\label{fig:fGC_fGC}
Marginalized hyper-posteriors of $\fGtoMain$ for all 18 universes.
The plot setting is the same as in Fig.~\ref{fig:firstPeaks}.
}
\end{figure}

The partial correlation between $\nF$ and $\nIII$ manifests itself in two other interesting ways.
First, we observe an increase in the uncertainty of $\fGtoMain$ when Pop~III mergers are present.
In Fig.~\ref{fig:fGC_fGC}, we show violin plots for the marginalized $\fGtoMain$ posteriors at different true $\fGtoMain$ values.
The purple and yellow violins correspond to the universes with and without Pop~III binaries, respectively. While the true values lie inside the 95\% credible interval in all cases, the uncertainties increase by ${\sim}10\%$ for the universes with Pop~III binaries.

\begin{figure}[h]
\includegraphics[width=0.9\columnwidth]{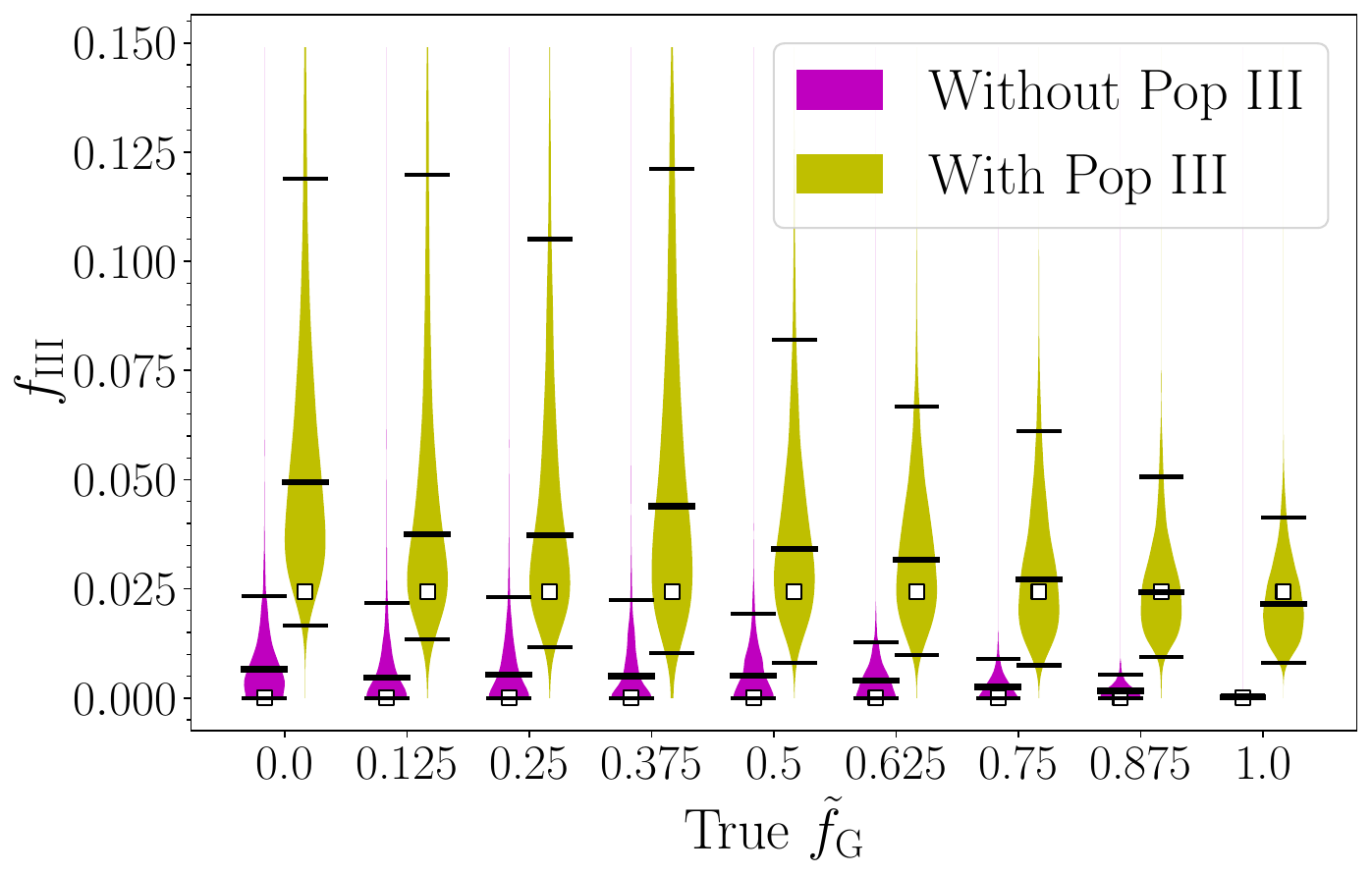}
\caption{\label{fig:fIII_fGC}
Marginalized hyper-posteriors of $\fIII$ for all 18 universes.
The plot setting is the same as in Fig.~\ref{fig:firstPeaks}.
}
\end{figure}

Second, the uncertainty in $\fIII$ decreases as $\fGtoMain$ increases.
Figure~\ref{fig:fIII_fGC} is a plot similar to Fig.~\ref{fig:fGC_fGC}, but showing the marginalized $\fIII$ posteriors in different universes.
When Pop~III binaries are present, the uncertainty stays roughly constant $\Delta \fIII \sim 0.1$ up to $\fGtoMain\lesssim0.5$, after which it drops gradually to $\Delta \fIII \sim 0.05$ at $\fGtoMain=1$.
A similar trend is observed when no Pop~III binaries exist: the uncertainty drops from $\sim2.5\%$ to $\lesssim 0.1\%$.
This is because $\nF$ has a longer tail in the high redshift $z\gtrsim 8$, and is easier to be confused with a small excess of Pop~III population.

\section{Discussions and Conclusions}\label{sec:discuss}

In this paper, we have shown that observations made by a network of 3G detectors can be used to infer the properties of different BBH populations using redshift-only information.
The larger horizon of a 3G detector network allows accessing thousands of BBHs per month up to $z\sim 15$, which is necessary to resolve the excess of high-redshift $(z\gtrsim8)$ BBHs originated from Pop~III stars. We consider $\sim16000$ binaries, roughly corresponding to two months of data, and multiple values of the branching ratio between binary formation in galactic fields and globular clusters. For every value of this branching ratio, we analyzed both a case where a few hundred Pop~III BBHs are present, and one where they do not exist.

First, we consider a hierarchical inference approach based on a nonparametric reconstruction of the total volumetric merger rate density $\nz$. We look for local peaks in the reconstructed total merger rate, and extract limited but useful information about the high-redshift population.
By requiring that a possible high-redshift peak has an amplitude of at least $1/10$ of the low-redshift peak, we find evidence for the presence of an high-redshift peak when Pop~III binaries are included, and constrain its position to be in the range $7\leq z_h\leq15$ for various mixing fractions between field and cluster binaries. Using the same approach, we rule out the existence of secondary-peak structure if there were no Pop~III binaries. This minimally modeled measurement of the position of an high-redshift peak (or lack thereof) in the total binary merger rate might, with some model, be translated into measurement or an upper limit on the abundance of Pop~III stars.
With a similar approach, we are able to measure the position of a low-redshift peak, which might be used to investigate typical time delays between start formation and mergers.

Then, we considered a modeled analysis where a phenomenological model exists for each of the three subpopulations, which are characterized by a set of unknown hyper-parameters, measured from the data together with the (unknown) branching ratios.
Among the most remarkable results, we found that irrespective of the true value of the relative abundances of field and cluster binaries, the Pop~III fraction can be constrained to be $\fIII\gtrsim0.01$ $(\fIII\lesssim0.02)$ at 95\% credibility for the universes that have (do not have) Pop~III binaries. In both cases, the branching ratio between field and cluster binaries can be measured with better than $\sim100\%$ uncertainty (95\% credible interval).
The precision on the measurement of the Pop~III population mainly depends on the morphology of the merger rate densities of the dominating channels in the high-redshift region.
If the dominating channels predict a shallower declining slope at high redshift, an eventual contribution to the high-redshift merger rate from the Pop~III population is less distinctive, introducing correlations with the dominating channels.

Some studies suggest that Pop~III population might contribute to a non-negligible fraction of the merger rate in the local Universe~\citep{Hartwig:2016nde,Liu:2020ufc}, or even a secondary peak in the low redshift $z\sim 2$ due to different formation scenarios~\citep{Kinugawa:2020ego,Liu:2020lmi}. If this additional low redshift peak exists, it will then make the Pop~III subpopulation more distinguishable, while degrading the measurements of branching ratios, owing to extra degeneracies in the low redshift regime.

We emphasize that our analysis is only assuming two months worth of data.
A back-of-the envelope calculation assuming that the statistical uncertainty shrinks like $1/\sqrt{N}$ would imply a factor of ${\sim}5$ improvement over the results we present here, after 5 years of data.
In that scenario, the phenomenological approach could identify a fraction of Pop~III mergers as small as $\fIII\sim0.5\%$.
However, as stressed multiple times in this work, the phenomenological inference requires reliable models for the merger rate of the various formation channels, and will yield results which are as good as the models. On the other hand, the model independent approach, though intrinsically less precise, has the attractive feature of not requiring any specific modeling of the underlying subpopulation.
In this analysis, we used a three-detectors network. A smaller network would lead to worse redshift measurements for individual sources, and hence yield worse statistical uncertainties than what reported in this paper. However, this can be compensated by a longer observation time.

In this work, we have only considered three subpopulations: the galactic field
and cluster binaries, as well as high-redshift Pop~III binaries. As mentioned above, many other channels have been proposed and can plausibly contribute a sizable fraction of the total rate.
Our analysis can be trivially extended to include these and other subpopulations, at the price of increasing computational cost, correlations, and potentially degrading the measurement of some of the parameters.
On the other hand, most of these different channels predict distinctive features in the BBHs they produce, beside their redshift distribution, for example masses, spins and eccentricity~\citep{Vitale:2015tea,OShaughnessy:2016nny,Dominik:2012kk,Dominik:2013tma,Dominik:2014yma,deMink:2015yea,Belczynski:2016obo,Mapelli:2019bnp,Breivik:2019lmt,Stevenson:2017tfq,Antonini:2020xnd,Santoliquido:2020axb,Rodriguez:2015oxa,Rodriguez:2016kxx,Rodriguez:2018rmd,DiCarlo:2019pmf,Kremer:2020wtp,Bartos:2016dgn,Yi:2019rwo,Yang:2019cbr,Yang:2020lhq,Grobner:2020drr,Tagawa:2019osr,Tagawa:2020qll,Tagawa:2020dxe,Samsing:2020tda,Raidal:2017mfl,Raidal:2018bbj,Biscoveanu:2020are}. Including these features can enhance the precision of multi-population inference, help fighting correlations, and improve the understanding of each formation channel.
We leave this extension of multi-dimensional BBH parameters in the 3G era as a future work.

\acknowledgments
The authors would like to thank Emanuele Berti, Hsin-Yu Chen, Carl Haster, and the LVK's rate and population working group for fruitful discussions and comments.
The authors also thank Katarina Martinovic, Carole Perigois and Tania Regimbau for cross-checking the validity of the Pop~III phenomenological model.
K.N. and S.V. acknowledge support of the National Science Foundation through the NSF award PHY-1836814.
K.N. and S.V. are members of the LIGO Laboratory.
W.M.F.\ is funded by the Center for Computational Astrophysics at the Flatiron Institute, which is supported by the Simons Foundation.
LIGO was constructed by the California Institute of Technology and Massachusetts Institute of Technology with funding from the National Science Foundation and operates under cooperative agreement PHY-1764464.
This paper carries LIGO document number LIGO-P2000540 and CE document number CE-P2000007.

\appendix
\section{Statistical models}\label{sec:stats}
Here, we briefly review the main statistical tool used in the analysis, i.e. the hierarchical Bayesian inference framework~\citep{Farr:2013yna,Mandel:2018mve,Thrane:2018qnx,Wysocki:2018mpo,Vitale:2020aaz}.
We model the production mechanism of BBHs as an inhomogeneous Poisson process whose differential merger rate in the detector frame~\footnote{$R$ and $R^d$ measure mergers per unit time. The clocks used to measure the time interval, the detector's or the ones comoving with the sources, determine the frame in which the rate is calculated.} is given by
\begin{align}
\dNdzDetFrame(z\mid \pmb{\Lambda}, \NdotDet) = \frac{1}{1+z} \dNdzSrcFrame(z\mid \pmb{\Lambda}, \Ndot),
\end{align}
where $\dNdz$ is the differential merger rate in the comoving frame characterized by the ``shape parameters'' $\pmb{\Lambda}$ and an overall normalization factor given by the total merger rate in the comoving frame $\Ndot = \int \dNdzSrcFrame dz$.
The factor $1/(1+z)$ accounts for the cosmological time dilation effect so that the total merger rate in the detector frame is
\begin{align}
\NdotDet = \int\dNdzDetFrame dz = \int \frac{1}{1+z} \dNdzSrcFrame dz.
\end{align}
The vector of shape parameters $\pmb{\Lambda}$ contains the quantities which are used to model the underlying physical populations (see Sec.~\ref{sec:injection}).

One can write the hyper-posterior of the population parameters $\pmb{\Lambda}$ and $\NdotDet$ given a set of $\Nobs$ observations $\pmb{d}\equiv \{ d_i \}_{i=1}^{\Nobs}$ as:

\begin{align}\label{eq:hyperpos}
& \quad p\left( \pmb{\Lambda}, \NdotDet \mid \pmb{d}\,\right) \nn \\
&\propto \left[\prod_{i=1}^{\Nobs} T^d \int dz_i \, p\left( d_i \mid z_i \right) \dNdzDetFrame \left(z_i \mid \pmb{\Lambda}, \NdotDet \right)\right]  e^{-\NdotDet T^d} \,\pi\left(\pmb{\Lambda}, \NdotDet \right) \nn\\
&\simeq \left[\prod_{i=1}^{\Nobs} T^d \frac{1}{M_i} \sum_{j=1}^{M_i} \dNdzDetFrame \left(z_{ij} \mid \pmb{\Lambda}, \NdotDet \right)\right] e^{-\NdotDet T^d} \, \pi\left(\pmb{\Lambda}, \NdotDet \right).
\end{align}

In going from the second to the last line of Eq.~\eqref{eq:hyperpos}, we have approximated the integrals with discrete sums. For the $i$-th source, this amounts to calculating an average of the merger rate evaluated at the $M_i$ points $\{z_{ij}\}_{j=1}^{M_i}$ drawn from the likelihood $p(d_i \mid z_i)$ of the $i$-th source.
In the third line, $\pi(\pmb{\Lambda}, \NdotDet)$ is the hyper-prior, and $T^d$ is the experiment duration in the detector frame.

We will find more useful to quote the \emph{volumetric} merger rate density, defined for the k$-th$ subpopulation as $$\dot{n}_k(z)\equiv d\Ndot_k/dV_c,$$ where $dV_c/dz$ is the differential comoving volume.
Then, the merger rate history $d\NdotDet_k/dz$ in the detector frame as a function of redshift for the $k$-th subpopulation is:
\begin{align}
\dNdzDetFramekth(z \mid \pmb{\Lambda}_k, \NdotDet_k) = \frac{1}{1+z} \frac{dV_c}{dz} \dot{n}_k(z \mid \pmb{\Lambda}_k, \Ndot_k),
\end{align}
where the subscript $k$ denotes the relevant quantities of the $k$-th subpopulation.

The overall merger rate can be expressed as the sum of the individual merger rates of all $P$ subpopulations ($P=3$ in our analysis), i.e.
\begin{align}\label{eq:dNdzTotal}
\dNdzDetFrame(z \mid \pmb{\Lambda}, \pmb{R}^d) = \sum_k^P \dNdzDetFramekth(z \mid \pmb{\Lambda}_k, \NdotDet_k),
\end{align}
where the vectors $\pmb{\Lambda}$ and $\pmb{R}^d$ contains all $\pmb{\Lambda}_k$'s and $\NdotDet_k$'s, respectively.
Since $\sum_k^P \Ndot_k=\Ndot$ and $\sum_k^P \NdotDet_k = \NdotDet$, we may rewrite Eq.~\eqref{eq:dNdzTotal} in terms of the branching ratios in the detector frame, i.e. the fraction of sources in each subpopulation, $f_k \equiv \NdotDet_k/\NdotDet$,
\begin{align}\label{eq:phenom}
\dNdzDetFrame (z \mid \pmb{\Lambda}, \NdotDet) &= \NdotDet \sum_k^P f_k p^d_k(z \mid \pmb{\Lambda}_k),
\end{align}
where $p^d_k(z \mid \pmb{\Lambda}_k)$ is the normalized merger rate of the $k$-th population in the detector frame, and the $f_k$'s are subject to the constraint $\sum_k^P f_k=1$.
Since we expect the fraction of Pop~III binaries to be small, it is more convenient to introduce the fraction of cluster binaries over the sum of field and cluster binaries, $\fGtoMain\equiv \fG/(\fG+\fF)$.

Therefore, for the parametrized analysis we model the merger rates of the three subpopulations in terms of several phenomenological parameters, which are treated as unknowns, together with the branching ratios (Sec.~\ref{sec:injection}). The total merger rate in Eq.~\ref{eq:hyperpos} is thus calculated by adding up the contribution of each channel. On the other hand, in the unmodeled approach, we measure the overall $\dNdzSrcFrame~{(z\mid \pmb{\Lambda}, \Ndot)}$ directly, without making any assumption about the individual subpopulations that might be contributing to it.

More details on the two approaches can be found in Appendices~\ref{sec:gpr}~and~\ref{sec:phenom}, whereas the hyper-priors are described in the Appendix~\ref{app:hyperprior}.
Throughout the study, we assume Planck 15 cosmology~\citep{Ade:2015xua}.

\section{Gaussian process regression}\label{sec:gpr}

This section provides details on the implementation of the GRP that we use to infer $\nz$ without assuming any specific functional form. We only require that $\dNdz$ is sufficiently smooth such that $\dNdz$ can be described by a piecewise-constant function over $W = \nbin$ redshift bins, which are uniformly distributed in linear space in the range $0\leq z \leq 15$.
The merger rate $\dNdz$ is thus written as
\begin{equation}
\frac{d\Ndot}{dz}
= \begin{cases}
\frac{\Delta \Ndot_1}{\Delta z_1} & 0 \leq z < z_1 \\
\ldots & \\
\frac{\Delta \Ndot_i}{\Delta z_i} & z_{i-1} \leq z < z_i \\
\ldots & \\
\frac{\Delta \Ndot_W}{\Delta z_W} & z_{W-1} \leq z < z_W
\end{cases},
\end{equation}
where $\Delta \Ndot_i$ is the merger rate in $i$-th redshift bin $\Delta z_i~{\equiv}~z_i - z_{i-1}=0.5$ so that $\sum_{i=1}^W (\Delta \Ndot_i)~{\equiv}~\Ndot$.
To make the GPR more efficient, we infer $\dNdz$ in natural-log space.
Then, we apply a squared-exponential Gaussian process prior on $X_i \equiv \ln{(\Delta \Ndot_i)}$, with a covariance kernel
\begin{align}
K_{ij} &\equiv \mathrm{Cov}\left( X_i, X_j \right) \nn \\
&= \sigma_{X}^2 \exp\left[-\frac{1}{2}\left(\frac{z_{i-1/2} - z_{j-1/2}}{l}\right)^2\right],
\end{align}
where $z_{i-1/2} = \frac{1}{2}\left( z_{i} - z_{i-1} \right)$ is the midpoint of the $i$-th redshift bin, $\sigma_X^2$ is the variance of $\{X_i\}$, and $l$ is the correlation length in redshift space.
The multivariate Gaussian process prior on the random variable vector $\pmb{X} \equiv \{\ln{(\Delta \Ndot_i)}\}$ with a mean vector $\pmb{\mu}_X$ and a covariance matrix $\mathbf{K}\equiv \{K_{ij}\}$ is then,
\begin{align}
\mathcal{G}(\pmb{X} \mid \pmb{\mu}_X, \sigma_X,l) \equiv \mathcal{N}\left[ \pmb{X} \mid \pmb{\mu}_X, \mathbf{K}\left(\sigma_X,l\right) \right].
\end{align}
The kernel $\mathbf{K}$ enforces the smoothness of $\dNdz$ on scales that are comparable to or larger than $l$, which may be much larger than the bin spacing if the data support it, and prevents from over-fitting when $W$ is large~\citep{Foreman-Mackey:2014}.
To further enhance the sampling efficiency, we utilize the Cholesky factorization to decompose $\mathbf{K}$ into a lower-triangular matrix $\mathbf{L}$ such that
\begin{align}
\pmb{X} \equiv \mu_X + \mathbf{L}(\sigma_X,l)\pmb{\eta}~
\end{align}
follows the same Gaussian process prior $\mathcal{G}$ by drawing $\pmb{\eta}$ from a multivariate standard normal distribution.
A common choice of $\pmb{\mu}_X$ is a constant mean vector $\mu_X \mathbf{1}$.
Since we know that $\dNdz$ has a strong dependence of the redshifted differential comoving volume $dV_c/dz/(1+z)$, we further impose the mean of the Gaussian process prior to be the natural log of $dV_c/dz/(1+z)$ (normalized to $\Nobs$ within the comoving volume $V_c$ and the observation time $T$) with a common shift $\Delta \mu_X$, i.e.
\begin{align}
\{\mu_{X,i}\} = \left\{ \ln\left(\frac{\Nobs}{V_cT}\int_{z_{i-1}}^{z_i}\frac{1}{1+z} \frac{dV_c}{dz} dz\right) + \Delta\mu_X \right\}~,
\end{align}
and we treat $\Delta \mu_X$, which is a single variable, as an additional parameter to include any possible fluctuation in the normalization $\Ndot$
We then obtain $\nz$ from $\dNdz$ divided by the differential volume in each bin.
Altogther, the $W+3=33$ hyper-parameters in the Gaussian process regression are thus
\begin{align}
\pmb{\Lambda}_{\rm GPR}=\left(\{\eta_i\}_{i=1}^{W}, \Delta\mu_X, \sigma_X, l  \right).
\end{align}

\section{Phenomenological models}\label{sec:phenom}

Directly modeling the volumetric merger rate $\nz$ of a subpopulation as a function of astrophysical quantities, such as various distributions of initial stellar mass, mass/radius of star clusters, BH natal kicks, or stellar metallicity, requires detailed stellar evolution or N-body simulations which are computationally expensive.
To facilitate our analysis, we model the three formation channels phenomenologically.
For field, we follow the Madau-Dickinson functional form~\footnote{Note that while we use the same \emph{form} for the equation, we do \emph{not} assume that the numerical coefficients are the same of the standard Madau-Dickinson SFR}:
\begin{align}\label{eq:field}
\nF(z \mid \alphaF, \betaF, \CF) \propto \frac{(1 + z)^{\alphaF}}{1 + \left(\frac{1+z}{\CF}\right)^{\betaF}},
\end{align}
where $\alphaF$, $\betaF$ and $\CF$ are unknown parameters that characterize the upward slope at~$z\lesssim \CF-1$, the downward slope at~$z\gtrsim \CF - 1$, and the peak location of the volumetric merger rate density, respectively.

For cluster binaries, we describe the volumetric merger rate as a log-normal distribution in cosmic time, which we treat as a function of redshift:
\begin{align}\label{eq:cluster}
\nG(z \mid \muG, \sigmaG, \tG) \propto \text{LogNorm}(t(z)-\tG \mid \muG, \sigmaG),
\end{align}
where $t(z)$ is the cosmic time as a function of redshift, and $\text{LogNorm}$ is the standard lognormal distribution of the argument $t-\tG$ parameterized by $\muG$ and $\sigmaG$.
The additional parameter, $\tG$, is a reference time that mark the birth of the first cluster binaries.

For Pop~III, we use the following functional form:

\begin{align}\label{eq:pop3}
\nIII(z \mid \aIII, \bIII, \zIII) \propto \frac{e^{\aIII(z-\zIII)}}{\bIII+\aIII e^{(\aIII+\bIII)(z-\zIII)}},
\end{align}
where $\aIII$, $\bIII$ and $\zIII$ characterize the upward slope at~$z<\zIII$, the downward slope at~$z>\zIII$, and the peak location of the volumetric merger rate density, respectively.

We have verified that these three phenomenological models can fit well the data from population synthesis analysis~\citep{Belczynski:2016obo,Rodriguez:2018rmd,Belczynski:2016ieo} for values of their arguments given in Eqs.~\eqref{eq:field},~\eqref{eq:cluster}~and~\eqref{eq:pop3}.

We define the branching ratio between field and cluster binaries, $\fGtoMain$ implicitly through the equations:

\begin{align}
\fG &\equiv \fGtoMain\left(1-\fIII\right),\\
\fF &\equiv \left(1-\fGtoMain\right)\left(1-\fIII\right),
\end{align}
where $\{f_k\}$ are the original fractions of Eq.~\eqref{eq:phenom}.

Therefore, there are a total of 12 hyper-parameters in the phenomenological model:
\begin{align}
\pmb{\Lambda}_{\rm PM}=\left( \alphaF, \betaF, \CF, \muG, \sigmaG, \tG, \aIII, \bIII, \zIII, \fGtoMain, \fIII, \NdotDet \right).
\end{align}

\section{Simulation details}\label{sec:injection}

In this section we describe how we prepare the simulated universes that will be analyzed with the methods described in the previous section.

First, we need to choose reference (i.e. ``true'') merger rate densities that will be used to generate the redshift of the BBHs.
We do so by means of the phenomenological curves in Eqs.~\eqref{eq:field},~\eqref{eq:cluster} and~\eqref{eq:pop3}.
As described in Appendix~\ref{sec:phenom}, these curves describe the morphology of each volumetric merger rate density, and are parametrized by, e.g., the rising slope at low redshift, the declining slope at high redshift, and the redshift at which the merger rate peaks. The phenomenological curves are obtained by fitting Eqs.~\eqref{eq:field},~\eqref{eq:cluster} and~\eqref{eq:pop3} to the simulation results available in the literature~\citep{Belczynski:2016obo,Rodriguez:2018rmd,Belczynski:2016ieo}. Specfically, we take the model-averaged simulation results of \citet{Belczynski:2016obo} and \citet{Rodriguez:2018rmd} for field and cluster binaries, respectively, as well as ``FS1'' model's result of ~\citet{Belczynski:2016ieo} for Pop~III binaries.
We stress that our phenomenological fits include the effect of the time delay from binary formation to merger, as well as the impact of stellar metallicity on the binary evolution specified in the population synthesis analyses. In particular, this implies a quite remarkable fact: the same Madau-Dickinson function can be used both to fit the star formation rate (this is its normal use) and the merger rate density it implies, for different values of its parameters. 

We use the following numbers as the ``true'' values of each curve parameters, when preparing our sources:

\begin{align}
\left( \hat{\alpha}_{\rm F}, \hat{\beta}_{\rm F}, \hat{C}_{\rm F} \right) &= (2.57, 5.83, 3.36), \nn \\
\left( \hat{\mu}_{\rm G}, \hat{\sigma}_{\rm G}, \hat{t}_{\rm G} \right) &= (1.63, 0.96, 0.66\mathrm{\,Gyr}), \nn \\
\left( \hat{a}_{\rm III}, \hat{b}_{\rm III}, \hat{z}_{\rm III} \right) &= (0.66, 0.3, 11.6).\nn
\end{align}

While this fixes the true shape of the merger rate for each subpopulation, we still need to fix the amplitudes.
We use 9 different values of relative the merger rate between cluster and field binaries, $\fGtoMain$, equally spaced in the range from 0 to 1.
For each value of $\fGtoMain$, we consider two universes: one with and one without Pop III binaries.

Following the latest GWTC-2 result, we fix the local volumetric merger rate density to $\dot{n}(0) = 25~\mathrm{Gpc}^{-3}\mathrm{yr}^{-1}$~\citep{GWTC2rate}.
This yields $\NdotDet_{\rm F}+\NdotDet_{\rm G} \approx 8000$ per month.
In the universes with Pop~III binaries, the Pop~III fraction $\fIII\approx 0.024$ is chosen to generate an additional $200$ sources per month coming from the Pop~III channel. This number is chosen such that the peak merger rate density of Pop~III binaries is ${\sim}10$ times smaller than the peak of the dominating channels.
The nominal value of this peak, $\nIII(\zIII) \approx 20 ~\mathrm{Gpc}^{-3}\mathrm{yr}^{-1}$, is consistent with the comparison of ``FS1'' model to the Pop I/II field binaries in~\cite{Belczynski:2016obo}.
We stress that the current predictions for the Pop~III merger rate span a few orders of magnitudes, in either direction, relative to the one we are using~\citep{Belczynski:2016obo,Hartwig:2016nde,Kinugawa:2014zha,Kinugawa:2015nla}.
This is because the formation efficiency of Pop~III binaries greatly depends on the initial mass function of the Pop~III stars and the distribution of the initial orbital separation (see~\cite{Belczynski:2016obo,Hartwig:2016nde}).
Since $\nIII$ declines rapidly at later cosmic time, it does not significantly contributes to $\dot{n}(0)$.

To ensure a reasonable measurement of luminosity distance, which typically requires three or more detectors, we choose a baseline detector network with one CE in Australia, one CE in the United States, and one ET in Europe~\citep{Vitale:2016aso,Vitale:2016icu,Vitale:2018nif}.
Generally speaking, not all BBHs within the detector horizon can be detected with SNRs over some thresholds, depending on their orientation and intrinsic parameters.
This results in a Malmquist bias~\citep{Mandel:2018mve,Vitale:2020aaz}. However, the efficiency only drops to 50\% at $z\sim20$ for a typical $30\Msun-30\Msun$ BBH.
Therefore, we limit our analysis to the redshift range $0\leq z \leq15$ and neglect selection effect.

Finally, we simulate a month of data in the following way.
We first draw the set of true redshifts, $\{z_{\mathrm{true}}\}$, from the ``true'' $\dNdz$ in each universe.
Then, for each $z_{\mathrm{true},i}$, we obtain the \emph{observed} redshift, $z_{\mathrm{obs},i}$, by drawing a random variable from a mock-up single-event likelihood conditional on $z_{\mathrm{true},i}$.
Following~\citet{Vitale:2018yhm}, we approximate the likelihood for redshift as a lognormal distribution conditional on $z_{\mathrm{true},i}$ with a standard deviation $\sigma_{\mathrm{LN},i}=0.017z_{\mathrm{true},i}$.
Finally, we draw 100 single-event likelihood samples conditional on $z_{\mathrm{obs},i}$ with the same $\sigma_{\mathrm{LN},i}$ calculated previously.
The second step is necessary in order to generate a scattering to the true value within the probable range of the single-event likelihood function.
Otherwise, the alignment of the true value and the mean of the likelihood introduces systematic bias in the analysis.

To summarize, we generate 18 simulated universes. In all universes, we generate 16000 (two-months worth data) BBHs from field and cluster binaries with a given branching ratio $\fGtoMain$, and add 400 more Pop~III observations for the 9 universes containing Pop~III binaries.

\section{Hyper-priors}\label{app:hyperprior}
The priors of all of the $\mathbf{\Lambda}_{\rm GPR}$ hyper-parameters are tabulated in Table.~\ref{tab:GPRpriors}.
\begin{table}[h]
\begin{ruledtabular}
\caption{Hyper-priors for the GPR model.
}
\label{tab:GPRpriors}
\begin{tabular}{cccc}
$\Lambda_{\mathrm{GPR},i}$ & Prior function & Prior parameters & Domain \\
\hline
$\eta_i$ & Normal & $(\mu_{\mathrm{N}},\sigma_{\mathrm{N}})=(0,1)$ & $(-\infty,+\infty)$\\
$\Delta\mu_X$ & Normal & $(\mu_{\mathrm{N}},\sigma_{\mathrm{N}})=(0,10)$ & $(-\infty,+\infty)$\\
$\sigma_X$ & Lognormal & $(\mu_{\mathrm{LN}},\sigma_{\mathrm{LN}})=(0,4)$ & $(0,+\infty)$\\
$l$ & Lognormal & $(\mu_{\mathrm{LN}},\sigma_{\mathrm{LN}})=\left(0,\frac{1}{2}\ln(10) \right)$ & $(0,+\infty)$\\
\end{tabular}
\end{ruledtabular}
\end{table}

The priors of all the $\mathbf{\Lambda}_{\rm PM}$ hyper-parameters are tabulated in Table.~\ref{tab:PMpriors}.
\begin{table}[h]
\begin{ruledtabular}
\caption{Hyper-priors for the phenomenological models.
}
\label{tab:PMpriors}
\begin{tabular}{cccc}
$\Lambda_{\mathrm{PM},i}$ & Prior function & Prior parameters & Domain \\
\hline
$\alphaF$ & Lognormal & $(\mu_{\mathrm{LN}},\sigma_{\mathrm{LN}})=(\hat{\alpha}_{\rm F},0.25)$ \footnote{$\mu_{\mathrm{LN}}$ and $\sigma_{\mathrm{LN}}$ are the mean and standard deviation of the lognormal distribution, respectively.}& $(0,10]$\\
$\betaF$ & Lognormal & $(\mu_{\mathrm{LN}},\sigma_{\mathrm{LN}})=(\hat{\beta}_{\rm F},0.25)$ & $(0,20]$\\
$\CF$ & Lognormal & $(\mu_{\mathrm{LN}},\sigma_{\mathrm{LN}})=(\hat{C}_{\rm F},0.25)$ & $(0,6]$\\
$\muG$ & Lognormal & $(\mu_{\mathrm{LN}},\sigma_{\mathrm{LN}})=(\hat{\mu}_{\rm G},0.25)$ & $(0,5]$\\
$\sigmaG$ & Lognormal & $(\mu_{\mathrm{LN}},\sigma_{\mathrm{LN}})=(\hat{\sigma}_{\rm G},0.25)$ & $(0,5]$\\
$\tG$ & Lognormal & $(\mu_{\mathrm{LN}},\sigma_{\mathrm{LN}})=(\hat{t}_{\rm G},0.25)$ & $(0,2.0]$\\
$\aIII$ & Lognormal & $(\mu_{\mathrm{LN}},\sigma_{\mathrm{LN}})=(\hat{a}_{\rm III},0.5)$ & $(0,2]$\\
$\bIII$ & Lognormal & $(\mu_{\mathrm{LN}},\sigma_{\mathrm{LN}})=(\hat{b}_{\rm III},1)$ & $(0,2]$\\
$\zIII$ & Normal & $(\mu_{\mathrm{N}},\sigma_{\mathrm{N}})=(\hat{z}_{\rm III},2)$ \footnote{$\mu_{\mathrm{N}}$ and $\sigma_{\mathrm{N}}$ are the mean and standard deviation of the normal distribution, respectively.} & $[8,20]$\\
$\fIII$ & Uniform & --- & $[0,0.5]$\\
$\fGtoMain$ & Uniform & --- & $[0,1.0]$\\
$\NdotDet$ & Half Cauchy & $\gamma_{C}=\Nobs$ \footnote{$\gamma_{C}$ is the scale parameter of the Cauchy distribution.}& $[0,+\infty]$\\
\end{tabular}
\end{ruledtabular}
\end{table}
The choice of lognormal priors is made to restrict each model always carrying a characteristic peak within the redshift range, rather than increasing or decreasing monotonically.
The gaussian prior and the domain of $\zIII$ ensure that the peak of $\nIII$ lies at high redshift $z\geq8$ and prevents $\nIII$ from mimicking the two dominating channels which have peaks at low redshift $z\lesssim3$.

\section{Detailed results from the modeled analysis}\label{app.Hyperpost}

In this section we report the population hyper-posteriors for the modeled analysis, i.e. the posteriors of the variables that parametrize the individual subpopulations described in Appendix~\ref{sec:phenom}, as well as the branching ratios.

\begin{figure}[htb]
\centering
\includegraphics[width=0.75\columnwidth]{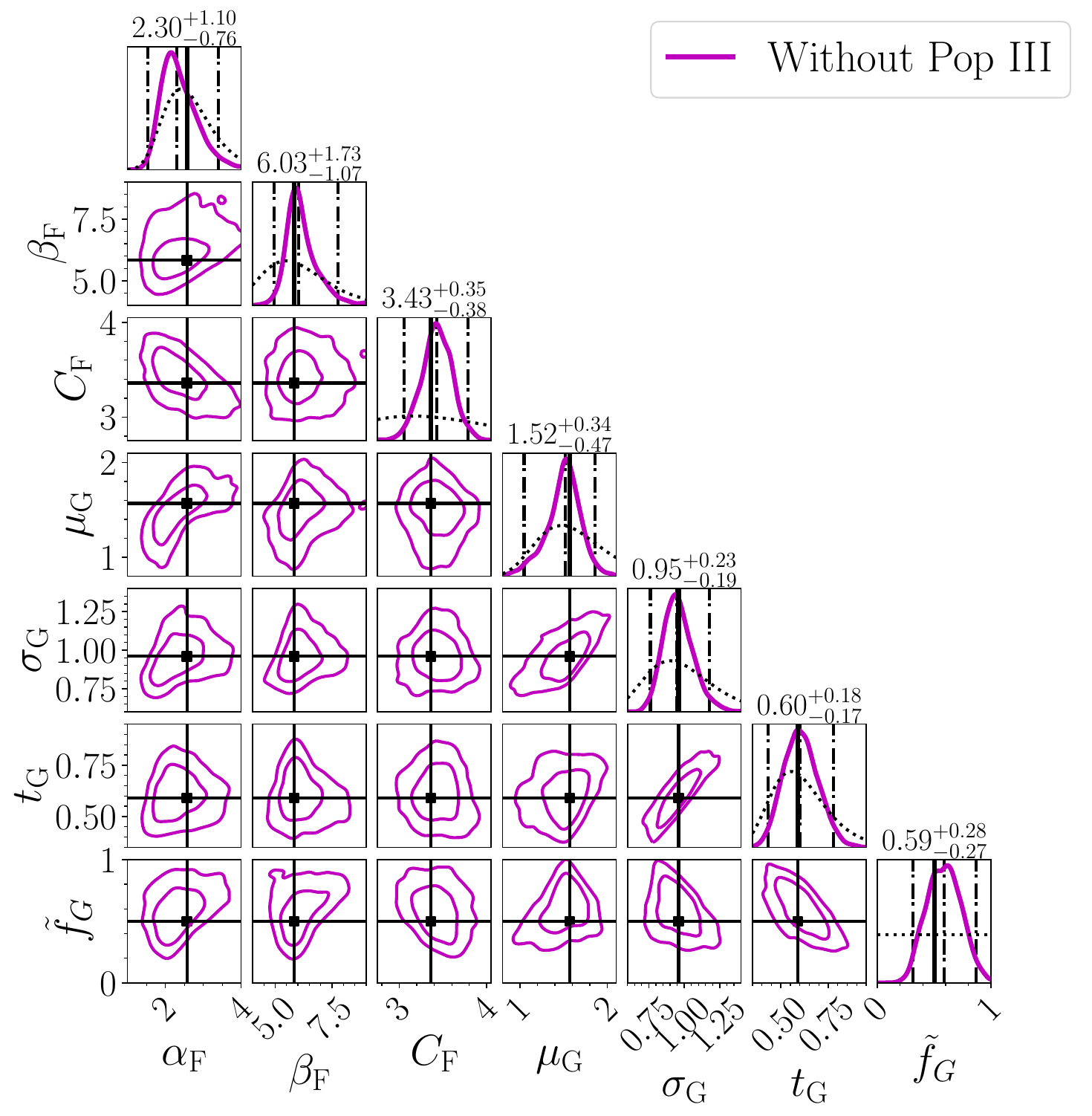}
\caption{\label{fig:hyperposFG}
Hyper-posterior of the dominating channels' hyperparameters, $(\alphaF,\betaF,\CF,\muG,\sigmaG,\tG,\fGtoMain)$, for the universe with $(\fGtoMain, \fIII)=(0.5,0)$ (the same of Fig.~\ref{fig:PMPosNull}).
For each marginalized 1D posterior (purple solid line in each diagonal slot), the left and right black dashed-dotted lines mark the 95\% highest posterior density credible interval, the middle black dashed-dotted line locates the median, and the black dotted line shows the prior.
The numerical values of median values and 95\% credible intervals are reported above the diagonal slots.
The off-diagonal slots show the marginalized 2D posteriors, with the contours representing the 68\% and 95\% credible intervals.
The black markers and solid lines indicate the true values, which lie within the 68\% credible interval.
The statistical behavior for the universe with Pop~III is very similar.
}
\end{figure}

First, we show the hyper-posterior of the dominating channels' parameters for the universe with $(\fGtoMain,\fIII)=(0.5,0)$ (the same of Fig.~\subref*{fig:PMPosNull}) in Fig.~\ref{fig:hyperposFG}.
Most of the shape parameters are measured with ${\sim}100\%$ uncertainty, e.g. $\fGtoMain=0.59_{-0.27}^{+0.28}$.
We note that there are a few interesting correlated pairs, such as $(\tG,\fGtoMain)$ and $(\tG,\sigmaG)$.
Since $\tG$ characterizes the starting time of the merging cluster binaries, an earlier $\tG$ shifts the peak of cluster population towards higher redshift where the cosmological volume is smaller.
Hence $\fGtoMain$ needs to be larger to keep the same number of cluster binaries.
On the other hand, a smaller $\tG$ tends to shift cluster population towards lower redshift.
Then a larger $\sigmaG$ is necessary to maintain a wide merger rate peak.

\begin{figure}[htb]
\includegraphics[width=0.5\columnwidth]{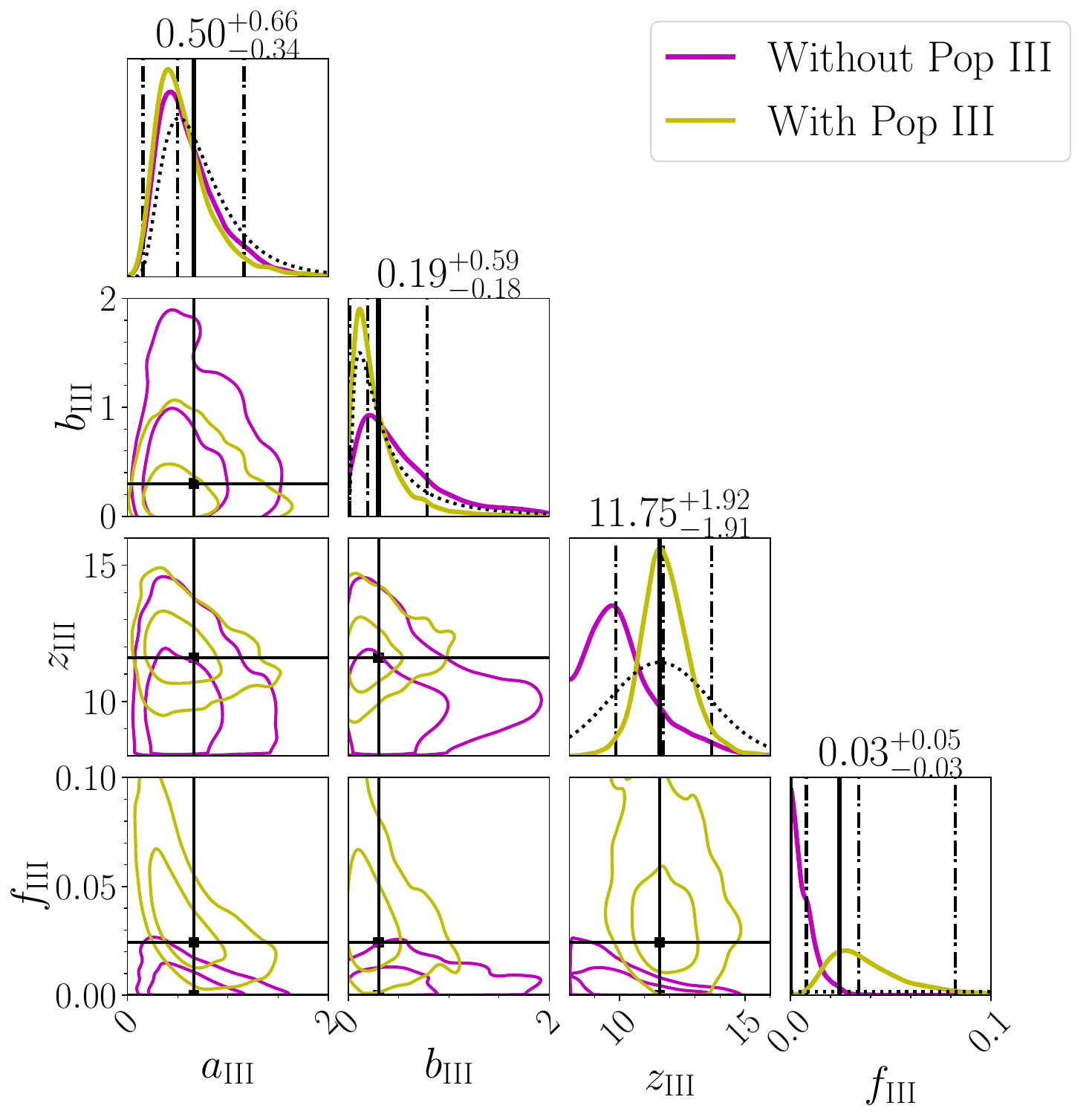}
\caption{\label{fig:hyperposPopIII}
Hyper-posterior of the Pop~III population's hyperparameters, $(\aIII,\bIII,\zIII,\fIII)$, for the universes $(\fGtoMain, \fIII)=(0.5,0)$ (purple) and $(\fGtoMain, \fIII)=(0.5,0.024)$ (yellow).
For each marginalized 1D posterior (purple solid line in each diagonal slot), the left and right black dashed-dotted lines mark the 95\% highest posterior density credible interval, the middle black dashed-dotted line locates the median, and the black dotted line shows the prior.
The numerical values of median values and 95\% credible intervals are reported above the diagonal slots.
The off-diagonal slots show the marginalized 2D posteriors, with the contours representing the 68\% and 95\% credible intervals.
The black markers and solid lines indicate the true values, which lie within the 68\% credible interval.
Only the medians and 95\% credible intervals for the universe with Pop~III binaries (yellow) are indicated by the dashed-dotted lines and reported above the diagonal slots.
}
\end{figure}

Next, we look at the hyper-posterior of the Pop~III merger rate parameters for the universe with $(\fGtoMain,\fIII)=(0.5,0)$ (the same of Fig.~\subref*{fig:PMPosNull}), as shown by the purple contours in Fig.~\ref{fig:hyperposPopIII}.
We are able to constrain $\fIII \lesssim 0.02$.
Both $\aIII$ and $\bIII$ are very close to their priors, which make sense since $\fIII$ is small which implies no information about $\nIII$ can be gained. On the other hand, the position of the peak, $\zIII$, is shifted towards lower values relative to its prior to suppress the contribution from $\nIII$ to the high-redshift total merger rate, which in this universe is entirely determined by the field binaries.

For comparison, we also report the results for the universe with $\fIII=0.024$ (this is the same as in Fig.~\ref{fig:PMPosPopIII}).
In Fig.~\ref{fig:hyperposPopIII}, the yellow contours show the hyper-posterior of the $\nIII$'s parameters.
While the shape parameters of $\aIII$ and $\bIII$ are still not well constrained, the peak $\zIII=11.75^{+1.92}_{-1.91}$ is measured with ${\sim}30\%$ relative uncertainty.
Importantly, $\fIII=0$, i.e., the absence of a Pop~III channel, is excluded from the 95\% credible interval of the marginalized $\fIII$ posterior.
This provides strong evidence of the existence of Pop~III binaries in our simulated data.

\clearpage
\bibliography{reference}

\end{document}